\documentstyle[12pt]{article}
\begin{document}
\title{ Nonlinear Physics: Integrability, Chaos and Beyond}
\author {\\M. Lakshmanan\\Centre for Nonlinear Dynamics\\
Department of Physics\\Bharathidasan University\\Tiruchirapalli 620 024\\
India}
\maketitle
\begin{abstract}
Integrability and chaos are two of the main concepts associated with
nonlinear physical systems which have revolutionized our understanding
of them. Highly stable exponentially localized solitons are often
associated with many of the important integrable nonlinear systems while
motions which are sensitively dependent on initial conditions are
associated with chaotic systems. Besides dramatically raising our
perception of many natural phenomena, these concepts are opening up new
vistas of applications and unfolding technologies: Optical soliton based
information technology, magnetoelectronics, controlling and
synchronization of chaos and secure communications, to name a few. These
developments have raised further new interesting questions and
potentialities. We present a particular view of some of the challenging
problems and payoffs ahead in the next few decades by tracing the early
historical events, summarizing the revolutionary era of 1950-70 when
many important new ideas including solitons and chaos were realized and
reviewing the current status. Important open problems both at the basic
and applied levels are discussed.
\end{abstract}
\newpage
\section {Introduction}
Fifty years is a long period in modern science and it is very hazardous 
to visualize how a given subject will grow and blossom in a period of five
decades.  Particularly this is so for an area like Nonlinear Science, whose
growth is quite recent and phenomenal, and even the most immediate 
ramifications of its exciting developments have not yet been fully understood
and utilized.  Even then it will be fruitful and relevant to diagress the 
past history, evaluate the present status and visualize the future course 
of directions and developments.

Of course if one asks the question how was nonlinear science 50 years ago, 
the answer is simply that no such field existed then and probably not many
had envisaged its emergence and potentialities.  It was the linear science
that generally ruled the scientific world, whether it was quantum 
mechanics or field theory, fluid mechanics or solid state physics or 
electronics or name any other field barring exceptions.  Linear physics
is associated with beautifully structured linear mathematics, including 
spectral theory, integral transforms, linear vector spaces, linear 
differential equations and so on.  There were only very few instances where
nonlinear systems in physics were considered to be important or relevant
or tractable.  One thought nonlinearities are essentially perturbations to 
exactly solvable linear systems, their effects could be analysed through 
well developed perturbation methods, statistical analysis and so on.  Of 
course there were isolated studies of nonlinear systems starting from the 
pendulum motion solvable by elliptic functions to Kepler problem and rigid 
body problem in classical mechanics, classification of nonlinear ordinary 
differential equations, geometrical theory of partial differential equations,  
and isolated problems in fluid mechanics and field 
theory.  Also gravitation theory is a patently nonlinear theory, where 
essentially special solutions were sought and obtained.  Thus except for 
such isolated studies, little systematic analysis of nonlinear systems on 
its own merit was carried out in general.

Yet there were enough warnings and forbadings by men of great forsight 
and vision during the past 100 years or so that a lot of new phenomena and 
insight are in store and are associated with nonlinear systems:
John Scott-Russel's observation of solitary wave and its 
remarkable stability properties, Sophia Kovalevskaya's analysis of rigid 
body problem based on singularity structure analysis, Poincare's analysis
on the sensitive dependence on initial conditions of nonlinear 
systems, van der Pol's observations of chaotic oscillations in 
electrical circuits, Einstein's insistence on the importance of nonlinear 
field theories and so on.  Each of these gems of ideas took a long time of 
gestation period before their implications could be understood.  One might 
say that essentially these ideas lie at the root of the modern development of 
nonlinear science.

The modern renaissance of nonlinear science in general and nonlinear physics 
in particular, of course, starts from the famous Fermi-Pasta-Ulam(FPU) 
experiments on the energy sharing phenomenon 
in nonlinear lattices in 1955.  The ensuing many faceted investigations 
on the one hand by the remarkable analysis of Zabusky and Kruskal in 1965 
led to the concept of solitons and ultimately to the development 
of integrable systems, while on the other hand led to 
the study of Hamiltonian 
chaos.  Parallel path breaking studies of Lorenz in 
1963 on the thermal convection equations ultimately led to the full power 
of chaotic dynamics.  During the same period the remarkable foresight of 
Skyrme has lead to the revival of interest in nonlinear field theories in 
particle physics.  In retrospect one might say that the period 1950-70 
is the golden age of nonlinear physics, revolutionizing our concept and 
outlook.  Some people even call it a third revolution in physics 
in this century, besides relativity and quantum mechanics.

The period 1970-96 is then a period of consolidation and enjoying the 
initial fruits of the earlier two era.  Much understanding in both the 
fields of integrable and chaotic systems have been achieved and their 
ramifications and applications have touched almost all fields of physics
during this period.

Though these developments are substantial, still many fundamental questions 
remain to be answered.  For example, when a given system is integrable and 
when it is nonintegrable and then chaotic is a tricky question which needs 
to be answered rigorously.  What are the natural excitations in higher 
dimensional systems?  Are there other types structures besides solitonic 
and chaotic structrues?  These are some of the crucial questions which must 
be answered soon. How are nonintegrable systems to be understood from the point 
of view of integrable systems and how can one obtain effective spatio
temporal patterns? What are the phenomena that are lying dormant to be
exploited for technological developments? What kind of technical
aspects one can develop in application oriented topics such as 
magnetoelectronics, optoelectronics, controlling and synchronization of
chaos and so on? These and other potential future problems in nonlinear
physics are also of considerable importance to be considered.

This article will start with a brief account of the historical 
perspectives.  Then it will discuss how based on these ideas the modern 
development of nonlinear science and the various topics in nonlinear physics 
it has led to developed.  Then the current status of each of these fields 
will be briefly reviewed.  In the last phase of the article, some of the 
outstanding problems in each of these areas will be discussed and finally 
the future perspective will be analysed.
\section { Early Visions of Nonlinear Science and their Physical  
Ramifications}

In this section, let us consider briefly some of the sailent features of the 
remarkable foresightedness of a few visionary scientists in realizing the 
importance of nonlinear phenomena.  Their discoveries or vision, though not 
fully appreciated or even drew derisive comments by contemporary scientists, 
withstood the test of time and each one formed a cornerstone of the modern 
concepts of nonlinear science and in particular nonlinear physics.
\subsection { Scott-Russel and his great wave of translation} 

John Scott-Russel, the Victorian entrepreneur and naval engineer for 
East India Company, was making investigations on the size of the ship-hull, 
weight, speed etc. in the Union canal connecting the cities of Edinburgh and 
Glasgow.  It is now a scientific folklore [Bullough, 1988] that during the 
course of these 
investigations he was careful enough to make the first scientific 
observation of a patently nonlinear phenomenon, namely the formation of a 
solitary wave which can propagate without change of speed and form and whose 
velocity is dependent on its amplitude.  This observation in the own words 
of Scott-Russel as he reported in the British Association Reports (1844) runs 
as follows.

``I was observing the motion of a boat which was rapidly drawn along a narrow 
channel by a pair of horses when the boat suddenly stopped-not so the mass 
of water in the channel which it had put in motion: it accumlated round 
the prow of the vessel in a state of violent agitation, then suddenly leaving 
it behind, rolled forward with great velocity, assuming the form of a large 
solitary elevation, a rounded smooth and well defined heaf of waterp, which 
continued its course along the channel apparently without change of form or 
diminution of speed.  I followed it on horseback, and overtook it still 
rolling on at a rate of some eight or nine miles an hour, preserving its 
orginal figure some thirty feet long and a foot to a foot and a half in 
height.  Its height gradually diminished and after a chase of one or two miles 
I lost in the windings of the channel.  Such, in the month of August 1834, 
was my first chance interview with the singular and beautiful phenomenon 
which I have called the Wave of Translation, a name which it now generally
bears: which I have since found to be an important element in almost every 
case of fluid resistance, and ascertained to be of the type of that great 
moving elevation of the sea, which, with the regularity of a planet, ascends 
our rivers and rolls along our shores.

To study minutely this phenomenon with a view to determining accurately its 
nature and laws, I have adopted other more convenient modes of products it 
than that which I just described, and have employed various methods of 
observation.  A description of these will probably assist me in conveying 
just conception of the nature of this wave.

Genesis of the Wave of the First Order ......"

Scott-Russel immediately realized the importance of his observations, went 
back to his laboratory and carried out a series of experiments during 
1834-40 and confirmed the permanent nature of the solitary wave.  
Scott-Russel also deduced a phenomenological relation between the velocity c 
and amplitude $\eta$,
$$c = \sqrt{g(h+\eta)}, \eqno(1)$$
where g is the acceleration due to gravity, h is the depth of the undisturbed 
water in the canal.  It is also remarkable that Russel had also realized 
even the collision properties of the solitary waves [Bullough,1988].  John 
Scott-Russel was obviously far ahead of his time in realizing the importance 
of such solitary waves of permanence in that his contemporaries such as Airy, 
Stokes and others refused to believe Scott-Russel's observations and 
explanations.  In fact Scott-Russel died as a disappointed man in 
that during his lifetime he could not make his fellow scientists to accept 
his findings. 

It took many more years before Boussinesq in 1872 and later on Korteweg and 
de Vries in 1895 could put Scott-Russel's observations in the proper 
perspective rigorously (For details see Bullough [1988]).
Starting from the basic equations of hydrodynamics and the fact
that there are two small parameters available in the problem, namely the ratio 
of the height(amplitude) of the water wave in the canal to the depth of the 
channel and the depth of the channel to the length of the canal (or solitary 
wave), and using fixed and free (nonlinear) boundary conditions, and by a 
systematic (asymptotic) expansions (see for example, Ablowitz and Clarkson, 
[1991]), Korteweg and de Vries showed that the Scott-Russel wave phenomenon 
can be described by a nonlinear wave equation for the amplitude $\eta(x,t)$.
Its modern version is the ubiquitous form
$$u_{t}+6uu_x+u_{xxx} = 0, \eqno (2)$$
where $u(x,t)$ is related to $\eta(x,t)$ linearly.
Eq.(2) admits solitary wave solutions of the form
$$ u(x,t) = 2k^2 sech^2 \left [k \left( x-4k^2 t-\delta \right ) \right] , \; 
k, \; \delta: constants.\eqno (3)$$
One can easily check that the empirical formula of Scott-Russel given in eq.(1) also 
naturally follows from eq.(3), showing the correctness of Russel's observations.

Unfortunately even after the Korteweg-de Vries analysis, which comprehensively 
showed the presence of an entirely new phenomenon, namely solitary waves of 
permanence being supported by nonlinear partial differential equations, 
little further interest seems to be have been shown by the scientific 
community in such a patently nonlinear phenomenon for another 70 years or so 
until Zabusky and Kruskal came across exactly at the same K-dV equation 
albeit in an entirely new physical situation, namely the propagation of 
waves in a nonlinear lattice-the famous Fermi-Pasta-Ulam(FPU) lattice.  It 
is now part of the scientific history that how the Zabusky-Kruskal work had led to the 
concept of soliton (and Russel's solitary wave is indeed a soliton) and how 
this concept is playing a paradigmic role in many branches of physics 
ranging from astronomy to particle physics, condensed matter, fluid dynamics, 
ferromagnetism, optical physics and so on to biological physics.  
\subsection { S.Kovalevskaya's work on integrability of dynamical systems 
and singularity structure analysis}

Sophia Kovalevskaya who migrated from Russia, and studied under Weisstrass,  
considered the problem of integrating the equations of motion of nonlinear 
dynamical systems.  Particularly she took up the problem of analysing for 
what parametric values the equations of motion of a rigid body (top) rotating 
about a fixed point [Kovalevskaya,1889] is completely integrable and 
analytic integrals of motion can be obtained.  This was a problem posed by 
the Paris Academy of Sciences for the Bordin Prize of 1888 and 
S.Kovalevskaya approached this problem in an entirely novel way, whose full 
ramifications are only realized in recent times with great potential 
applications for the future. 

Considering the dynamics of a rigid body with one fixed point under the 
influence of gravitation the equation of motion can be written as
$$A\frac{d\Omega_1}{dt} = (B-C)\Omega_2\Omega_3-\beta x_0+\gamma y_0,
\; \frac{d\alpha}{dt} = \beta \Omega_3-\gamma \Omega_2,\eqno(4a)$$
$$B\frac{d\Omega_2}{dt} = (C-A)\Omega_1\Omega_2-\gamma x_0+\alpha z_0,
\; \frac{d\beta}{dt} = \gamma \Omega_1-\alpha \Omega_3,\eqno(4b)$$
$$C\frac{d\Omega_3}{dt} = (A-B)\Omega_1\Omega_2-\alpha y_0+\beta x_0,
\; \frac{d\gamma}{dt} = \alpha \Omega_2-\beta \Omega_1,\eqno(4c)$$
for the components of the angular velocity vector $\overrightarrow{\Omega}$ 
and angular momentum $\vec I$
$$\overrightarrow{\Omega} = \sum\nolimits_{i=1}^3 \Omega_i \vec e_i,\; 
\vec I = A \Omega_1 \vec e_1 + B \Omega_2 \vec e_2 + C \Omega_3 \vec e_3,
\eqno(4d)$$  
with respect to the moving trihedral $\vec e_i, i=1,2,3$ fixed on the body.    
Here the vertical unit vector $\vec e$ and the centre of mass $\vec r_0$ are 
given by
$$\vec e = \alpha \vec e_1+\beta \vec e_2+\gamma \vec e_3, \; 
\vec r_0 = x_0 \vec e_1+y_0 \vec e_2+z_0 \vec e_3.\eqno(4e)$$
In order to identify the parametric choices for which the nonlinear dynamical 
system (4) becomes completely integrable Kovalevskaya used the novel idea 
that the solutions of integrable cases will be meromorphic (that is solutions 
will be free from movable critical points, namely movable branch points 
and essential singularities) in the complex time plane, just as in the case of 
equations satisfied by elliptic functions.  S.Kovalevskaya was far ahead of 
her time in realizing the connection between meromorphicity and integrability.  
It took almost a century before mathematical physicists could appreciate 
such an approach and even now the implications are not clearly understood.  

S.Kovalevskaya was probably motived by the works of R.Fuchs, who isolated 
that class of odes whose solutions are meromorphic from out of all first 
order differential equations of the form
$$\frac{dy}{dx} = F(x,y), \eqno(5)$$
where $F$ is analytic in $x$  and algebraic in y. It was shown that only the 
solution of Riccati equation
$$\frac{dy}{dx}+P_1(x)y+P_2(x)y^2 = P_3(x),\eqno(6)$$
is free from movable critical points.  Further the fact that elliptic 
functions are meromorphic might have lead Kovalevskaya to seek solutions of 
the form
$$x_i(t) = \sum\nolimits_{n=0}^\infty a_{i,n}(t-t_0)^{\displaystyle n-
\displaystyle p_i},\eqno(7)$$
for the dynamical variables in eq.(4) where $p_i$ is an integer, which is to 
be determined along with the coefficients $a_{i,n}$.

Kovalevskaya identified essentially four nontrivial parametric choices from 
the above analysis for which eqs.(4) are integrable, out of which three were 
already known:

$i) A = B = C$ (well known)

$ii) x_0 = y_0 = z_0$ (due to Euler)

$iii) x_0 = y_0 = 0, A = B$ (due to Lagrange)

$iv) y_0 = z_0 = 0, A = B = 2C$ (new)

For each of the four integrable cases, four independent involutive 
integrals of motion exist [Lakshmanan and Sahadevan, 1993].  For the first 
three cases, the solutions are expressible in terms of elliptic functions 
which are meromorphic, while for the Kovalevskaya's fourth case, the 
solutions are given by hyperelliptic functions, which are in this particular 
case again meromorphic [Kruskal and Clarkson, 1991].
Even after one century, these results stand the test of time and no further 
new integrable cases have been found, thereby proving 
Kovalevskaya's farsighted intuition.

Unfortunately the method was not pursued further by dynamical systems 
community until late 1970s, except for the mathematicians P.Painlev\'{e} and 
his coworkers, Gambier, Garnier and so on(during 1900-10).  The latter 
authors isolated those second order differential equations, whose solutions 
are free from movable critical points, of the form
$$\frac{d^2 y}{dx^2} = F(\frac{dy}{dx},y,x),\eqno(8)$$
where F is a rational function of $\frac{\displaystyle dy}{\displaystyle dx}$ 
and y, and analytic in x.  
Painlev\'{e} and coworkers showed that out of all possible equations (8), 
there are only fifty or so canonical types which have the property of their
solutions having no movable critical points, out of which 44 are solvable by 
elementary functions including elliptic functions and the remaining required 
new transcendental functions, the so called Painlev\'{e} transcendental 
fuctions.  For fuller details see Ablowitz and Clarkson [1991], Lakshmanan 
and Sahadevan [1993].

Again unfortunately, these important developments have not received much 
attention among the scientific community and very little work was done on the 
classification of higher order ordinary differential equations or partial 
differential equations, except for some isolated studies on third order 
odes by Bureau, Chazy and so on (see for example, Ablowitz and Clarkson, 
1991).  The full implications of these investigations have to again wait for 
many more decades until 1980s when the connection to soliton equations was 
established.

\subsection { Poincar\'{e}'s work on sensitive dependence on initial 
conditions}
Henri Poincar\'{e} (1854-1912), the pioneering mathematician, physicist and 
philosopher of the early part of this century, in his famous works on 
celestical mechanics was involved with the problem of stability of motion of 
dynamical systems, like the three body gravitational problem, and the problem 
of finding precise mathematical formulas for the dynamical history of complex 
systems.  In the course of these studies he was led to the notion of 
(Poincar\'{e}) surface of section and the concept of sensitive dependence of 
motions on initial conditions.

Poincar\'{e} concluded in his essay on Science and Method "It may happen  
that small differences in the initial conditions produce very great ones in 
the final phenomena.  A small error in the former will produce an enormous 
error in the latter.  Prediction becomes impossible".  Thus Poincar\'{e} 
conceived the notion of deterministic chaos and the associated motion which 
is sensitively dependent on initial conditions in nonlinear systems even as 
early as the beginning of this century (for more details see for example 
Brillouin[1964] and Holmes[1990]).  However it took several decades 
before Poincar\'{e}'s ideas could be understood for their full ramifications.

\subsection { van der Pol's investigations on coexisting multiperiodic 
solutions in forced nonlinear oscillators}
The coexistence of several multistable periodic solutions and non-periodic 
(highly unstable) solutions in forced nonlinear oscillators was realized as 
early as 1927 in their remarkable experimental study by van der Pol and van 
der Mark [1927].  They analysed essentially the circuit given in 
Fig.1 [Jackson, 1991].  It has a neon glow lamp Ne, a battery 
$E(\approx 200V)$, a resistor R(of several megaohms), and an applied emf
$E_0(\approx 10V)$.  In the absence of the emf the period of the system 
increases with increasing capacitance C.

There are essentially three important discoveries associated with the 
experiments of van der Pol and van der Mark. \\
\hspace{3cm} 1) Presence of more than 40 subharmonics $(\Omega/n)$ of the 
applied frequency, $\Omega$.\\
\hspace{3cm} 2) These subharmonics were found to be entrained over a limited 
range of C.  As C was further varied, the frequency changed discontinuously 
to another subharmonic(Fig.2).\\
\hspace{3cm} 3) Observation of bands of 'noise' in the regions of many 
transitions of the frequency, which was regarded as a 'subsidiary phenomena'.  
Also their figure clearly showed a hysterisis effect, hence a dynamical 
bistability in the system.

Thus van der Pol-van der Mark's observations were clearly the precursors to  
the to modern theory of chaotic nonlinear oscillators.  Again it took more than 
five or six decades to fully understand the implications of these far 
reaching discoveries.  However, one must also note that the experiments of 
van der Pol and van der Mark had historically lead to a number of important 
theoretical works to understand nonlinear phenomena in the underlying 
oscillator systems, notable among which is the work of Cartwright and 
Littlewood[1945] and of Levinson[1949].  The latter authors works, though 
abstract, were concerned with the analysis of the underlying dynamical 
equation to describe the current/voltage in Fig.1 in the form
$$\ddot{x}+k(\dot{x}^2-1)\dot{x}+x = fcos\Omega t. \hskip 10pt
(. = \frac{d}{dt})\eqno(9)$$
They noted the existence of the limit cycle solution to (6).  Ultimately 
all these works got perfected in the recent works in nonlinear dynamics(see 
for example, Lakshmanan and Murali, 1996).
\subsection{ Einstein's observations}

Starting from the days of electromagnestism there had been attempts to 
describe elementary particles in terms of localized solutions of nonlinear 
field equations.  Typical examples in this direction are Mie's theory of 
point particles and nonlinear electrodynamics of Born and Infeld (see for 
example, Schiff [1962]).  
It is interesting to note Einstein's view [Einstein, 1965] regarding classical 
nonlinear field theories.  He remarks:``Is it conceivable that a classical 
field theory permits one to understand the atomistic and quantum structure 
of reality? .... I believe that at present time nobody knows anything reliable 
about it ... We do not possess any method at all to derive systematically 
solutions that are free of singularities.  Approximate methods are of no 
avail since one never knows whether or not there exists to a particular 
approximate solution an exact solution free of singularities.  Only a 
significant progress in the mathematical methods can help here ...'

The above remark very clearly makes profound prediction of the relevance of 
nonlinear dynamics.

\section { The Revolution: Integrability and Chaos}
The period 1950-70 can be considered as the golden age of nonlinear physics 
when revolutionary discoveries were made on integrable and chaotic systems 
leading to the present advances.  Among these developments, the celebrated 
Fermi-Pasta-Ulam(FPU) experiments and the 
associated paradox [Fermi, Pasta \& Ulam, 1955] may 
be rightly considered as the harbinger of a new era in physics.  The various 
attempts to explain the FPU paradox ultimately resulted in the discovery of 
`solitons' in the KdV equation by Martin Kruskal and Norman Zabusky in 1965 
[Zabusky and Kruskal, 1965], simultaneously giving an integrable 
approximation resolution of the FPU paradox.

Interestingly, the concept of chaos was also found to be lurking behind the 
FPU experiments.  In fact the simplest FPU lattice can be mapped onto the 
celebrated Henon-Heiles system discovered in 1964 [Henon and Heiles, 1964] 
exhibiting the notion of chaos, namely sensitive dependence on initial 
conditions, when the nonlinearity is sufficiently large.  In the same period, 
the atmospheric scientist Lorenz discovered that a grossly reduced set of 
convection equations in the form of a 
set of three coupled first order ordinary nonlinear differential 
equations, namely Lorenz equations [Lorenz, 1963],
also show sensitive dependence on initial
conditions leading to dissipative chaos, thereby heralding the era
of chaotic dynamics.

The above first mentioned works on soliton systems were followed by the 
further discoveries of inverse scattering method and other soliton generating 
techniques for a large class of nonlinear dispersive systems in (1+1) 
dimensions and confirmed the fact that the KdV solitons are not fortuitous 
entities but form one of the most important basic coherent structures of 
nonlinear dynamics.  This has led to the stage for the application of 
solitons in diverse fields of physics.  
Similarly the path breaking discoveries of Henon and Heiles, and of Lorenz  
that certain 
nonlinear systems can exhibit sensitive dependence on initial conditions 
culminated in the development of possible routes to and characterization of 
chaos in the 1970s, leading to revolutionary implications in physics.

We will concisely discuss these developments in the following sections.

\subsection { The FPU paradox: A harbinger of revolutionary era of 
nonlinear physics}
In the early 1950s, Enrico Fermi, Stan Ulm and John Pasta were set to make 
use of the MANIAC-I at Los Alamos Laboratory on important problems in 
physics(for a detailed account see for example Ford, [1992]). In particular, 
Fermi felt that it will be highly instructive to integrate the equations of 
motion for judiciously chosen, one dimensional, harmonic chain of mass points, 
weakly perturbed by nonlinear forces.  The expectation was that the state 
of the chain as it evolves could not be accurately predicted after a finite 
time and it could form a simple model to test the various sophisticated 
questions related to irreversible statistical mechanics.  To begin with they 
intended to test the simplest and most widely believed assertions of 
equilibrium statistical mechanics such as equipartition of energy, ergodicity 
and the like.

The FPU model is essentially the one-dimensional chain of (N-1) moving mass 
points having the Hamiltonian
$$H = \sum\nolimits_{i=1}^{N-1} \frac{P_i^2}{2}
+\frac{1}{2}\sum\nolimits_{i=0}^{N-1} (Q_{i+1}-Q_i)^2
+\frac{\alpha}{3}\sum\nolimits_{i=0}^{N-1} (Q_{i+1}-Q_i)^3,\eqno(10)$$
where $Q_0=Q_N=0$ and $Q_i$ and $P_i$  are the coordinate and momentum of the 
ith particle, and $\alpha$ is a small nonlinearity parameter.  FPU had in 
addition considered quartic and broken linear couplings.

Now looking through a normal mode decomposition
$$A_l = \sqrt{\frac{2}{N}}\sum\nolimits_{k=1}^{N-1}Q_k \sin(\frac{kl\pi}{N}), 
\eqno(11)$$
one has essentially a system of independent harmonic oscillators weakly 
coupled by terms cubic in the normal mode positions, given by the Hamiltonian
$$H = \frac{1}{2}\sum(\dot{A}_k^2+\omega_k^2 A_k^2)+
\alpha \sum C_{klm}A_k A_l A_m, \eqno(12)$$
where $\omega_k = 2 \sin(\frac{\displaystyle k \displaystyle {\pi}}
{\displaystyle 2N})$ is the frequency of the kth 
normal mode and $\displaystyle C_{\displaystyle klm}$ are constants.  Then 
$E_k = \frac{1}{2}(\dot{A}_k^2+\omega_k^2A_k^2)$ 
is the energy of the kth normal mode and to a first 
approximation $H=\sum E_k$, when $\alpha$ is small.

How does $E_k$ change as a function of time when the weak nonlinear forces 
are present?  The results of FPU for the lattice (12) (or (10)) with N=32 
and $\alpha=\frac{1}{4}$ are given in Fig.3, with an initial shape at $t=0$ 
in the form of a half-sine wave given by 
$Q_k = \sin(\frac{\displaystyle {k\pi}}{\displaystyle 32})$ so that 
only the fundamental harmonic mode was excited with an amplitude $A_1=4$ and 
energy $E_1 = 0.077\cdots$.  During the time interval 
$0 \le t \le 16 $ in Fig.3, where t is measured in periods of the 
fundamental mode, modes 2,3,4, etc. sequentially begin to absorb energy 
from the initially dominant first mode as one would expect from a standard 
analysis.  After this, the pattern of energy sharing undergoes a dramatic 
change.  Energy is now exchanged primarily only among modes 1 through 6 with 
all higher modes getting very little energy.  In fact the motion of the 
anharmonic lattice is almost periodic and even perhaps quasiperiodic, with a 
recurrence period (FPU recurrence) at about $t=157$ fundamental periods.  The  
energy in the fundamental mode returns to within $3\%$ of its value at $t=0$.

FPU immediately realized that the above results are simply astounding.  First, 
they appear to violate the canons of statistical mechanics, which assert that 
the above type of nonlinear system should exhibit an approach to equilibrium 
with energy being shared equally among degrees of freedom.  But even more 
astonishing, they seem to invalidate Fermi's theorem regarding ergodicity in 
nonlinear systems.  Indeed, Fermi is said to have remarked that these results 
might be one of the most significant discoveries of his career.

Though the FPU results are truly path-breaking, it is curious to know that it 
took almost ten years for the matter to reach open literature that too just 
as part of Fermi's collected works [see Ford, 1992].  Originally a preprint 
was available in November 1955 as Los Alamos preprint 
LA-1940(7 November 1955), but then unfortunately Fermi died and the paper 
was never published.  This had a rather inhibitory effect as many people 
took the view that the results are too preliminary and it did not undergo 
peer review and perhaps does not warrant full attention.  Neverthless, the 
results became familiar through word of mouth and personal discussions and 
many serious efforts started being made to explain the FPU paradox.

\subsection { Integrable approximation: The Zabusky-Kruskal discovery 
of soliton}
Kruskal and Zabusky, through an asymptotic analysis, had sought a continuum 
approximation to FPU(see for example, Ablowitz and Clarkson, [1991]).  As the 
equation of motion of the anharmonic FPU lattice (10) can be written as
$$\ddot{Q}_k = (Q_{k+1}-2Q_k+Q_{k-1})[1+\alpha(Q_{k+1}-Q_{k-1})], 
\eqno(13)$$
in the lowest order continuous limit it takes the form
$$Q_{tt} = Q_{xx}+\varepsilon Q_x Q_{xx} = (1+\varepsilon Q_x)Q_{xx}.
\eqno(14)$$
When $\varepsilon=0$, eq.(14) is just the wave equation and when 
$\varepsilon \neq 0$, eq.(14) is hyperbolic and can develop shock.  Going 
over to the next order correction, under suitable asymptotic limit, Kruskal 
and Zabusky showed that the shock formation can be avoided with the addition 
of suitable dispersion so that a useful approximation to the FPU lattice 
results. Its specific form read 
$$Q_{tt} = Q_{xx}+\varepsilon Q_x Q_{xx}+\beta Q_{xxx}.\eqno(15)$$
Restricting attention to unidirectional waves, with the replacement of $x$ 
by $\sigma = x-t$, $t$ by $\tau = \varepsilon t$, $Q_x$ by  
$U = \frac{1}{2}Q_x = \frac{1}{2}Q_{\sigma}$ in eq.(15), and neglecting the 
terms proportional to $\varepsilon^2$, they obtained the celebrated Korteweg-
de Vries equation
$$U_{\tau}+UU_{\sigma}+\delta^2 U_{\sigma\sigma\sigma} = 0, \eqno(16)$$
which under rescaling can be recast in the standard form
$$u_t+6uu_x+u_{xxx}=0. \eqno(17)$$
Eq.(17) is nothing but the Korteweg-de Vries equation (1) derived to represent 
the Scott-Russel phenomenon as described in Sec.2.1, but now appearing in an 
entirely different physical context.

Of course eq.(17) admits solitary waves of the form (2), but Zabusky and 
Kruskal[1965] went on further to numerically integrate eq.(16)(and so (17)) 
using periodic boundary conditions and one cycle of a cosine as initial 
condition.  Much to their surprise, the initial cosine shape evolved into a 
finite number of relatively short pulses(Fig.4) that moved at distinct speeds 
about their periodic paths like runners on a track.  Upon collision, the 
pulses would exhibit a nonlinear superposition during overlap and then would 
emerge unchanged in shape or speed.

The almost-periodic behaviour of the FPU systems could now be understood at 
an especially clear and intuitive level.  The first full recurrence of the 
FPU motion occurs when all the pulses approximately overlap, generating a 
near return to the initial cosine shape.  The recurrence period calculated
by Zabusky and Kruskal closely approximates the actual FPU recurrence period, 
showing that KdV is a suitable long wavelength approximation of FPU system.  
Thus the Zabusky-Kruskal analysis provides an intuitively delightful 
interpretation of the FPU phenomenon, wherein the power of nonlinearity is 
made transparent.

But more interestingly the Zabusky-Kruskal experiments paved the way to 
provide insight into a much larger class of problems, turning the FPU 
paradox into a real discovery in physics (and mathematics), namely the 
invention of solitons, which are ubiquitous in nature.  The KdV equation 
itself has become a paradigm for an expanding class of completely integrable 
nonlinear differential equations of dispersive type admitting soliton 
solutions, and they possses Lax pair and solvable by inverse scattering 
transform procedure.  In particular the solitary waves of these systems 
under collision retain their shape and speed except for a phase shift, as 
demonstrated by Zabusky and Kruskal[1965] for the KdV equation.  Such 
solitary waves have been termed as solitons due to their particle-like 
properties by Zabusky and Kruskal.  A typical two soliton scattering 
property (for the KdV) is illustrated in Fig.5.

\subsection { Lax pair and inverse scattering formulation of KdV}
The remarkable stability properties of the soliton solutions of the KdV 
equation and the asymptotic form of the solutions in the form of N-number of 
solitons in the background of small amplitude dispersive waves had motivated 
Kruskal and coworkers to search for analytic methods to solve the initial 
value problem of the KdV equation.  In particular Gardner, Greene, Kruskal 
and Miura[1967] realized that the KdV equation is linearizable in the sense that 
it can be associated with two linear differential operators L and B, the so 
called Lax pairs.  Considering the time-independent Schr\"{o}dinger spectral 
problem
$$L \psi = \lambda \psi,\; L = -\frac{\partial^2}{\partial x^2}+u(x,t)
,\eqno(18)$$
where $\lambda$ is the eigenvalue parameter and $u$ is the unknown 
potential in which $t$ is a parameter and the associated time evolution 
equation for $\psi(x,t)$, 
$$ \psi_t = B\psi,\; B = -4\frac{\partial^3}{\partial x^3}
-6u\frac{\partial}{\partial x}-3u_x, \eqno(19)$$
then the compatibility of eqs.(18) and (19) with the condition that $\lambda$ 
is a constant in time leads to the Lax equation
$$L_t = [L,B] \Longleftrightarrow KdV. \eqno(20)$$

Thus given the initial condition $u(x,0)$ and analysing the linear equations 
(18) and (19), the initial value problem can be solved and all the numerical 
results of Zabusky and Kruskal discussed in the earlier sections can be 
obtained exactly.  For example the two soliton solution of the KdV can be 
obtained in the form
$$u(x,t) = 2(k_2^2-k_1^2) \frac{k_2^2 cosech^2 \gamma_2+k_1^2 sech^2 \gamma_1}
{(k_2 coth \gamma_2 - k_1 tanh \gamma_1)^2}, \eqno(21)$$
$$\gamma_1 = k_1 x-4k_1^3 t+\delta_1, \; \gamma_2 = k_2 x-4k_2^3 t+\delta_2,$$
where $k_1,k_2,\delta_1,\delta_2$ are constants, whose structure is exactly 
the same as given in Fig.5.  Similarly the explicit form of N-soliton 
solutions can also be obtained(see for example, Ablowitz and Clarkson, [1991]).  
KdV equation has also the remarkable property that it is a completely 
integrable infinite dimensional dynamical system in the sense that it 
possesses infinite number of involutive, functionally independent integrals 
of motion, which can be directly related to its linearizability property
(see Sec.4. below).

\subsection { Other soliton equations in (1+1) dimensions}
The KdV equation is ubiquitous in the sense it occurs in a large number of 
physical problems in areas as disparate as fluid dynamics, condensed matter 
physics, quantum field theory and astrophysics and so on.  Interestingly it is not 
only the KdV equation that admits solitons - there exists a large number of 
equally ubiquitous nonlinear dispersive wave equations of great physical 
importance which also admit soliton solutions, the major classes of which 
have been found during the decade following the work on KdV by Kruskal and 
his coworkers and the list is ever growing (for details, see for example 
Ablowitz and Clarkson, [1991]). Informations about some of the well known 
equations in (1+1) dimensions are given in Table I.

The most crucial aspects of soliton equations is that it is not only the KdV 
equation but also a large class of equally important systems such as the one 
given in Table I which become linearizable in terms of different Lax pairs 
or linear spectral problems.  For example, the Zakharov-Shabat(Z-S) and 
Ablowitz-Kaup-Newell-Segur(AKNS) matrix spectral problem 
$$V_x = MV,\eqno(22)$$
with the corresponding time evolution
$$V_t = NV,\eqno(23)$$
gives rise to the compatibility equation
$$M_t-N_x+[M,N] = 0,\eqno(24)$$
which is equivalent to the Lax equation (20).  Different choices of M and N 
lead to different nonlinear evolution equations.  The modified KdV, the 
nonlinear Schr\"{o}dinger, the sine-Gordon and the Heisenberg ferromagnetic 
spin equations are some of the important evolution equations which are 
linearizable in the above sense.  The linearized forms are also given in 
Table I.

Once the linearization is effected in the above sense that for a given nonlinear 
system $u_t = k(u)$, where $k(u)$ is a nonlinear functional of u and its 
spatial derivatives, the Cauchy initial value problem corresponding to the 
boundary condition $ u \to 0$ as $x\to\pm\infty$ can be 
solved by a three step process shown schematically 
in Fig.6.  This process which is known as the Inverse Scattering 
Transform(IST) procedure may be considered as the nonlinear analogue of the 
Fourier transform method applicable to linear dispersive systems.  The method 
consists of the following steps:\\
(i) \underline { Direct scattering analysis}: An analysis of the linear 
eigenvalue problem with the initial condition $u(x,0)$ as the potential is 
carried out to obtain the scattering data $S(0)$.  For example for the KdV, 
$$S(0) = \left\{k_n(0), \; n=1,2,\ldots, N, C_n(0),
R(k,0),\; -\infty < k <\infty \right\}.\eqno(24)$$
Here $N$ is the number of bound state eigenvalues $k_n$, $C_n(0)$ are the 
normalization constants of the bound state eigenfunctions, and $R(k,0)$ is the 
reflection coefficient for the scattering states.\\
(ii) \underline { Time Evolution}: Using the asymptotic form of the time 
evolution equation for the eigenfunctions, the time evolution of the scattering 
data $S(t)$ is determined. \\
(iii) \underline { Inverse Scattering}: The set of Gelfand-Levitan-
Marchenko linear integral equations corresponding to the scattering data 
$S(t)$ is constructed and solved.  The resulting solution consists typically 
of $N$ number of localized, exponentially decaying(soliton) solutions 
asymptotically ($ t \rightarrow \pm \infty $).

Thus once a given nonlinear evolution equation is fitted into the Lax pair 
and inverse scattering formalism, its Cauchy initial value problem can be 
solved, soliton solution obtained and completely integrability proved.  For 
a more general list of such integrable equations, we refer the reader to 
Ablowitz \& Clarkson [1991].

\subsection { The FPU lattice, Henon-Heiles system and transition to chaos}
The fact that the FPU lattice given by eq.(10) did not give rise to the 
expected statistical behaviour and that the integrable KdV approximation 
gives satisfactory explanation of the recurrence behaviour in the long 
wavelength limit made many people to wonder whether the FPU lattice itself 
is a completely integrable dynamical system.  However this doubt was soon 
dispelled by the following fact[Ford, 1992].  Consider a three-particle FPU 
system 
having periodic boundary conditions which is governed by the Hamiltonian
$$H = \sum \nolimits_{k=1}^{3}\frac{P_k^2}{2}+
\frac{1}{2} \sum \nolimits_{k=1}^{3}(Q_{k+1}-Q_k)^2
+\frac{\alpha}{3}\sum \nolimits_{k=1}^{3} (Q_{k+1}-Q_k)^3,\eqno(25)$$
where $Q_4=Q_1$.  After introduction of a canonical change of variables to 
harmonic normal mode coordinates $(\xi_k,\eta_k)$, the Hamiltonian (25) 
becomes
$$H = \frac{1}{2}(\eta_1^2+\eta_2^2+\eta_3^2)+\frac{3}{2}(\xi_2^2+\xi_3^2)     
+\frac{3 \alpha}{\sqrt 2}(\xi_2 \xi_3^2-\frac{1}{3}\xi_2^3).\eqno(26)$$
Now transforming to the centre of mass frame and setting 
$t=\frac{\displaystyle \tau}
{\displaystyle \sqrt 3},\; \eta_2 = \frac{\displaystyle \sqrt 2}
{\displaystyle \alpha}\displaystyle q_2$, and 
$\eta_3 = \frac{\displaystyle \sqrt 2}{\displaystyle \alpha} 
\displaystyle q_1$, the Hamiltonian (26) can be rewritten as
$$H = \frac{1}{2}(p_1^2+p_2^2+q_1^2+q_2^2)+q_1^2 q_2-
\frac{1}{3}q_3^2.\eqno(27)$$
But the Hamiltonian (27), canonically equivalent to the three-particle FPU 
lattice, is the celebrated Henon-Heiles system originally introduced 
by astronomers Henon and Heiles [1964] during the same revoltionary era which
shows the other facet of nonliner dynamical systems, namely chaos in 
deterministic systems.

Henon and Heiles were  interested in determining whether a third integral 
existed which constrained the motion of a star in a galaxy which had an
axis of symmetry.  Such a system has three degrees of freedom and two 
known isolating integrals of motion which are the energy and one component
of angular momentum.  It was long thought that such systems do not have a
third isolating integral because none had been found analytically.  However,
the nonexistence of a third integral implies that the dispersion of 
velocities of stellar objects in the direction of the galactic center is 
the same as that perpendicular to the galactic plane.  What was observed,
however was a 2:1 ratio in these dispersions.  Henon and Heiles [1964] 
constructed exactly the Hamiltonian (27) to model the essential features 
of the problem and studied it numerically by solving the equation of motion,
$$\frac{dq_i}{dt} = p_i,\; i=1,2,\eqno(28a)$$
$$\frac{dp_1}{dt}=-q_1 -q_1 q_2,\; 
\frac{dp_2}{dt} = -q_2 -2 q_1^2-2q_2^2.\eqno(28b)$$

A sketch of their results is shown in Fig.7 in the form of a Poincar\'{e} 
surface
of section.  At low energy (Fig.7a) there appears to be a third integral, 
at least to the accuracy of these figures (enlargement of the region around 
the hyperbolic fixed points 
would show a scatter of points).  As the energy is increased (equivalent to 
the increase of the effect of the nonlinear terms as may be seen from a 
scaling argument)  the third integral appears to be destroyed in the 
neighborhood of the hyperbolic fixed points (Fig.7b).  At still higher 
energies (Fig.7c), the second isolating integral appears to have been totally 
destroyed.  The scattered points in the Henon-Heiles plots correspond to a 
single trajectory which is chaotic.  Such trajectories are chaotic in that 
they have "sensitive dependence on intial conditions".  This is an entirely 
new kind of structure, originally envisaged by Poincar\'{e} (cf sec.2.3)  but 
which
is transparent in a simple Hamiltonian system now.  Additionally, this shows
that the original FPU system is not integrable and it can develop complex
motion if the strength of the nonlinearity in eq. (10) is increased.  (For
fuller details see for example Reichl [1992]).  And with this the era of chaos 
in Hamiltonian system has started in right earnest in physics!
\subsection {The Lorenz system and dissipative chaos}

Edward Lorenz, an atmospheric scientist and meterologist, who was 
interested in the long term behaviour of the atmospheric weather, reduced 
[Lorenz, 1963] the bare essentials of the underlying dynamical equations
to a system of three first order nonlinear coupled differential equations
$$\dot{x} = -\sigma(x-y),$$
$$\dot{y} = rx - y - xz,$$
$$\dot{z} = -bz + xy. \eqno(29)$$

The above equations essentially represent a model of two-dimensional 
convection in a horizontal layer of fluid heated from below.  In eq.(29)
x represents the velocity and y, z the temperature of the fluid at each
instant, and r,$\; \sigma,\;$ b are positive parameters determined by the 
heating of the layer of the fluid, the physical properties of the fluid,
and the height of the layer.

The general understanding in dynamics, particularly in weather prediction,
is strongly influenced by Laplace dictum of
complete predictability of solving Newton's equations of motion.  This 
means small deviations in the input or at any stage of the calculation
will lead to small uncertainities only in the output also.  When Lorenz
numerically integrated the equations of motion (29), which constitutes 
a dissipative system as the phase space volume shrinks,
$$\overrightarrow{\nabla}.\overrightarrow V = \frac{\partial \dot x }
{\partial x}+\frac{\partial \dot y }{\partial y}+
\frac{\partial \dot z }{\partial z}\ = -(\sigma+b+1)< 0,\eqno(30)$$
he discovered the now ubiquitous chaotic behaviour, the so called 
butterfly effect:  small deviations can grow exponentially fast in a
finite time - in the popular language as small as the effect of a 
butterfly fluttering its wings somewhere in the Amazons can lead to a
tornados in Texas in a few days time.

The physical consequence is that the long term weather prediction becomes
almost impracticable and geometrically in the phase space dynamical systems
can admit limiting motions which are extremely sensitively dependent on 
initial conditions (strange attractors).  Combined with Henon-Heiles
invention of similar effect in Hamiltonian systems, the concept of chaos
has now come to stay in dynamics.

Of course it is now well known that the Lorenz system admits several kinds
of dynamical motions: equalibrium points, limit cycles of different periods,
chaotic motions, strange attractors and so on.  The known results for a 
typical choice of the parameters 
to illustrate them and the form of chaotic attractors are given in 
Figs.8.

\subsection{The Skyrme's model}
During the same revolutionary era, the distinguished English physicist
T. H. R. Skyrme proposed a model of baryons as topological solitons in a
series of papers during the period 1955-63 (for details see for example
Chados etal, [1993]). Unfortunately the model was mostly ignored by particle
physics community until 1980s, when it was realized that the Skyrme model
could be considered as the possible low energy limit of quantum
chromodynamics. Since then the model has received enormous attention and
reverence.

In the Skyrme model, the nucleus is considered to be a classical,
electrically neutral imcompressible (mesonic fluid), which occupies a region
with radius $R$. The nucleons are immersed into this mesonic fluid, which
saturates them, while freely moving inside it. Further, the mesonic fields
could take their values on $S^3$ as the field manifold so that conserved
topological charge, which can be interpreted as the baryon number, is
associated with it. The Lagrangian suggested by Skyrme was of the form
$$L=-\frac{1}{4 \lambda^2}Tr \overrightarrow {L_{\mu}}^2
+\frac{\epsilon^2}{16}Tr \left [\overrightarrow {L_{\mu}}, \;
\overrightarrow {L_{\nu}} \right ]^2, \eqno(31)$$
where the chiral currents $\overrightarrow {L_{\nu}}$ are vector fields
defined on the $S^3$ manifold with values on the $s(u,2)$ algebra.
Suggesting a hedgehog ansatz, Skyrme suggested [Skyrme, 1962] that the
model could describe a stable extended particle with a unit topological
charge and all finite dynamical characteristics, which should be quantized
as fermions. Though the response was belated, the particle physics community
has realized the importance of Skyrme's model as one of the most fundamental
theories.

In the above discussions in the present section 3, we have tried to give a 
very personal bird's overview of the golden era of nonlinear dynamics when 
both the concepts of solitons and chaos were born and the role of the FPU
paradox, highlighting the analysis of the KdV equation by Zabusky and Kruskal leading 
to the concept of solitons and the discovery of sensitive dependence
on initial conditions in Henon-Heiles and Lorenz systems.  These pioneering
works will then obviously have far reaching consequences in physics.  The
remaining period of this century is then essentially a period of further
understanding of these developments and consolidation of these results and                         
application of these concepts in various physical systems.  These are taken
up in the next section.

\section {The Modern Era of Nonlinear Physics: Coherent and Chaotic
Structures and Their Applications}

The period starting from 1970s has seen an exponential growth of research 
in nonlinear science, particularly in nonlinear physics and mathematics.  
Correspondingly novel areas of applications have been identified to utilize 
both the concepts of coherent and chaotic structures.  The identification 
of more than a hundred completely integrable soliton systems in (1+1) 
dimensions and some possible extensions to (2+1) dimensions to identify 
exponentially localized solutions and the development of various analytic 
and algebraic techniques to isolate and investigate soliton systems are some 
of the important developments regarding integrable nonlinear systems in this 
period.  Similarly, identification of various routes to chaos, 
characterization of chaotic attractors and identification of numerous 
chaotic dynamical systems in nature are few of the main progress in chaotic 
dynamics during the last three decades.  There has been tremendous amount 
of applications of these ideas and potential technologies are unfolding due 
to these studies.  We may mention a few of them: magnetoelectronics using 
nonlinear magnetic excitations, soliton propagation in optical fibres as a 
means of lossless propagation revolutionizing  information technology and 
various opto-electronic devices in nonlinear optics, controlling aspects of 
chaos with its various ramifications including weather forecasting, 
synchronization of chaos and secure communications
are some of the important payoffs
realized during this current era.  We will briefly discuss these 
developments in this section.
\subsection { Soliton equations and techniques}

As noted in the previous section, it is now almost 30 years since the soliton 
was invented by Zabusky and Kruskal in their numerical experiments on KdV, 
followed by the invention of the inverse scattering transform method by 
Gardner, Greene, Kruskal and Miura[1967] to solve the Cauchy initial value 
problem of it for vanishing boundary conditions (at infinity).  This has 
ultimately led to the notion of complete integrability of infinite 
dimensional soliton systems, the list of which is ever expanding.  
Consequently the field has grown from strength to strength (see, for example, 
Ablowitz and Segur [1981], Ablowitz and Clarkson [1991], Fokas and Zakharov 
[1992], Lakshmanan [1988,1993,1995]).  A great number of increasingly 
sophisticated mathematical concepts from linear operator theory, complex 
analysis, differential geometry, Lie algebra, graph theory, algebraic 
geometry, and so on are being ascribed to soliton phenomenon, while new 
physical, engineering and biological appplications, where the soliton concept 
is found to play a crucial role, appear all the time (Lakshmanan [1995]).

There are two broad theoretical appraoches available to deal with the soliton 
bearing nonlinear evolution equations, namely (i) analytic and (ii) algebraic 
methods, though overlapping and strong interconnections exist between them 
(Fig.9).  Analytic approaches include the IST method its generalization 
(namely, the d-bar appproach) for solving the Cauchy initial value problem, 
and other soliton generating techniques such as the Hirota bilinearization 
method(Matsuno, [1984]), operator dressing method(Novikov et al, [1984]), 
B\"{a}cklund transformation method (Rogers and Shadwick, [1982]), direct 
linearizing transform method (Ablowitz and Clarkson, [1991]), apart from 
Painlev\'{e} analysis(Weiss et al, [1983]) to test integrability.  On the other 
hand, complete integrability aspects of soliton systems including the 
existence of infinite number of integrals of motion can be associated with 
the generalized Lie-B\"{a}cklund symmetries and the associated group 
theoretic, Lie algebraic, bihamiltonian structure and differential geometric 
properties(Asano and Kato, [1991]; Dickey, [1991]; Magnano and Magri, [1991]).  
Such complete integrability aspects can be further generalized to the area 
of quantum integrable systems, exactly solvable statistical models and so on 
(Wadati et al, [1989]) through Yang-Baxter relations and the quantum inverse 
scattering method.  Also the study of perturbation of soliton systems, often 
leading to spatiotemporal complexity, is of great physical interest in 
condensed matter physics, fluid dynamics, nonlinear optics, liquid crystals 
and so on (Sanchez and Vazquez, [1991]; Hasegawa, [1989]; Lam and Prost, [1991]).  
Thus one finds the study of soliton-bearing systems is of fundamental 
importance in several branches of physics and natural sciences.  We will now 
briefly mention some basic ideas of these various aspects.
\subsection {Direct methods}

The inverse scattering formalism described earlier in Sec.3.4 is quite 
sophisticated, although elegant.  Often one would like to have simpler 
analytic methods to obtain explicit N-soliton solutions.  Several invaluable 
direct methods have been developed in the literature for this purpose during 
the last three decades.  Among them, the Hirota's bilinearization method  
and the B\"{a}cklund transformation method have played very crucial roles in 
the development of the field as they help one to quickly obtain soliton 
solutions even when the IST formalism is not yet available for a particular 
evolution equation.  Besides, they have deep algebraic and geometric  set-ups,  
giving special significance to soliton systems.  Apart from these methods, 
the prolongation structure method of Wahlquist \& Estabrook[1975], the 
dressing method [Zakharov \& Shabat, 1974], the Darboux method [Matveev \& 
Salle, 1991] and the direct linearizing transform method [Santini et al., 
1984] are some of the other important techniques available for soliton 
solutions.  In this section we give the salient features of the Hirota method 
and the B\"{a}cklund transformation procedures 
only.  For the other methods, which have close connections with the above 
two and IST, the reader may refer to the references cited.
\subsection { Hirota's bilinearization method}
The essence of the method [Hirota, 1980; Ablowitz \& Segur, 1981; Matsuno, 
1984] is (i) to convert the given nonlinear evolution equation into what is 
known as bilinear forms, wherein each term is of degree 2 in the dependent 
variables, and (ii) to derive a formal power series solution which turns out 
to be the soliton solution for all soliton bearing NLEEs.

As an example, we again consider the KdV equation (17).  Under the 
transformation
$$ u = 2\frac{\partial^2}{\partial x^2}logF,\eqno(32a)$$
Eq.(17) takes the bilinear form (after 
integration and setting the integration constant to zero)
$$F_{xt}F-F_{x}F_{t}+F_{xxxx}F-4F_{xxx}F_{x}+3F_{xx}^2 = 0.\eqno(32b)$$
It is advantageous to introduce the so-called Hirota's bilinear operator
$$D_x^m D_t^n(a.b) = (\partial _x-\partial _{x'})^m 
(\partial _t-\partial _{t'})^n a(x,t)b(x',t')|_{x=x' \atop t=t'}, \eqno(33)$$
so that Eq.(32) takes the notationally simpler form
$$(D_x D_t+4D_x^4)F.F = 0. \eqno(34)$$
The properties of the bilinear operator can be easily worked out.  Samples:
$D_m^m a.1 = \partial _x^m a$, $D_m^m a.b = (-1)^m D_x^m b.a$,  
$D_x^m a.a = 0$, m odd, and so on.  Using such properties the calculations  
can be simplified considerably.

Now expanding F in a formal power series in $\varepsilon$ [Ablowitz \& Segur,
 1981],
$$F = 1+\varepsilon f^{(1)}+ \varepsilon ^2 f^{(2)}+\cdots,\eqno(35a)$$
where
$$f^{(1)} = \sum\nolimits_{i=1}^N e^{\eta_i}, \eta_i=k_i x+\omega_i t
+\eta_i^{(0)}\eqno(35b) $$
and $k_i, \omega_i, \eta_i^{(0)}$ are constants, the N-solitons of KdV can be 
obtained.  To see this, we substitute (34) in (33), equate each power of 
$\varepsilon $ separately to zero and obtain to 0($\varepsilon ^3$)
$$ 0(1) : 0 = 0, \eqno(36a)$$
$$ 0(\varepsilon) : 2(\partial_x \partial_t+\partial_x^4)f^{(1)} = 
0, \eqno(36b)$$
$$ 0(\varepsilon ^2) : 2(\partial_x \partial_t+\partial_x^4)f^{(2)} = 
-(D_x D_t+D_x^4)f^{(1)}.f^{(1)}, \eqno(36c)$$
$$ 0(\varepsilon ^3) : 2(\partial_x \partial_t+\partial_x^4)f^{(3)} = 
-2(D_x D_t+D_x^4)f^{(1)}.f^{(2)}, \eqno(36d)$$
The procedure is then to use (35b) in (36b) and succesively solve the 
remaining equations.  In practice, one finds the solution for N=1,2,3 and 
then hypothesize it for arbitrary N which is to be proved by induction.

For example, for N=1, $f^{(1)} = e^{\eta_1}$ and from (36b), 
$\omega _1 = -k_1^3$ and 
$(\partial_x \partial_t + \partial_x^4)f^{(2)} = 0$, so that $f^{(2)} = 0$ and  
$f^{(i)} = 0, i > 2$.  Thus the solution of (34) can be written as
$$ F_1 = 1+e^{\eta_1}, \; \eta_1 = k_1 x -k_1^3 t +\eta_1^{(0)}.\eqno(37)$$
Making use of (31), it is straightforward to obtain the 1-soliton solution 
(3) of the KdV equation.

Similarly for the case N=2, 
$$ f^{(1)} = e^{\eta_1} +e^{\eta_2}, \; \eta_i = k_i x -k_i^3 t +\eta_1^{(0)}, 
i=1,2$$
so that the solution of (34) becomes
$$ F_2 = 1+e^{\eta_1}+e^{\eta_2}+e^{\eta_1+\eta_2+A_{12}},\eqno(38)$$
where $A_{12}$ is constant.  Again this leads to the 2-soliton solution (21) of 
the KdV discussed in Sec.2 by using the transformation (31).  In an analogous 
fashion, one can proceed to find the N-soliton solution also.  We also note 
that with the solution of the Hirota equation in a form such as (38), it is 
quite easy to understand the elastic nature of the soliton interaction 
discussed in Sec.2.

As noted above, all the other known soliton equations can also be bilinearized 
and the soliton solution obtained through the Hirota method.  Even though 
the Hirota method appears to be merely a device to generate 
soliton solutions, recent investigations reveal deeper meaning to the role of 
bilinear theory; new, beautiful group-theoretic and geometric connections
can be ascribed to this method and one finds such integrable equations live in 
a phase-space which is an infinite dimensional Kac-Moody Lie algebra [Date 
et al., 1983].
\subsection { B\"{a}cklund transformations } 
B\"{a}cklund transformations (BTs) are essentially a set of relations 
involving the solutions of differential equations.  They arose originally 
in the theories of differential geometry and differential equations as a 
generalization of contact transformations (see, for example, Rogers \& 
Shadwick [1982]).  A classical example is the Cauchy-Riemann conditions, 
$u_x = v_y$ and $v_x = -u_y$, so that they are a BT from the Laplace equation 
into itself, as both u and v satisfy the Laplace equation.  A BT is 
essentially defined as a pair of partial differential equations involving 
two dependent variables and their derivatives which together imply that each 
one of the dependent variables satisfies separately a partial differential 
equation.  Thus, for example, the transformation
$$ v_x  = F(u,v,u_x,u_t,x,t),\\$$
$$ v_t  = G(u,v,u_x,u_t,x,t),\eqno(39)$$
will imply that u and v satisfy pdes of the operational form
$$ P(u) = 0, \; Q(v) = 0. \eqno(40)$$
If P and Q are of the same form, then (40) is an auto-BT.

Thus for the sine-Gordon equation the original transformation of B\"{a}cklund 
derived in 1880 is 
$$\big ( \frac{u+v}{2} \big)_x = k sin \big ( \frac{u-v}{2} \big),$$
$$\big ( \frac{u-v}{2} \big)_t = \frac{1}{k} sin \big ( \frac{u+v}{2} \big),
\eqno(41)$$
where k is a parameter, so that u and v satisfy the sine-Gordon equation, 
$\phi_{xt} = sin \phi$.  Similarly for the KdV, one can write down the 
BT as 
$$ w_x +w_x' = 2k-\frac{1}{2}(w-w')^2, \eqno(42a)$$
$$ w_t+w_t' = (w-w')(w_{xx}-w_{xx}')-2(u^2+uu'+u'^2),\eqno(42b)$$
where $w_x =u$ and $w'_x =u'$, so that u and $u'$ satisfy (17).  Similarly 
BTs can be worked out for all other soliton evolution equations also.  For the 
various methods available to obtain BTs, see for example Miura[1976], 
Rogers \& Shadwick[1982].

BTs can in priniciple be integrated to generate higher-order solitons, 
though in practice they are difficult to solve in general.  The usual way to 
get around this difficulty is to utilize the permutability property of the 
BT, which states that if we make two successive BTs from a given initial 
solution $u_0$, we will end up with the same final solution no matter what 
sequential order of the two BTs we take.  In other words, if $k_1$ and $k_2$ 
represent the parameters of the two BTs and if 
$$ u_0\stackrel{k_1}{\longrightarrow} u_1 \stackrel{k_2}{\longrightarrow} u_{12}, \;
u_0 \stackrel{k_2}{\longrightarrow} u_2 \stackrel{k_1}{\longrightarrow} u_{21},$$
then $u_{12} = u_{21}$.  From such permutability one can derive a nonlinear 
superposition formula, expressing $u_{12}$ algebraically in terms of $u_0$, 
$u_1$ and $u_2$.  For example for the sG equation, from (40) the following 
superposition law can be derived:
$$ u_{12} = u_0+4 tan^{-1} \left [ \frac{k_2 +k_1}{k_2 -k_1}
tan \left (\frac{u_1 - u_2}{4} \right ) \right ]. \eqno(43)$$
The superposition law can be repeatedly applied to construct soliton solutions 
of higher and higher order and one can make symbolic computation packages 
such as MAPLE effectively for this purpose.  

Another effective method to derive explicit BTs is to use the so-called 
dressing method [Zakharov \& Shabat, 1974].  This method requires solving a 
certain factorization problem in order to generate new solutions from a 
given input solution.  In the case of (1+1) dimensions, this factorization 
problem takes the form of a Riemann problem in the complex eigenvalue plane 
of the associated linear system.  

\subsection { Complete integrability of soliton systems }

The IST solvable soliton equations may be considered to be completely 
integrable infinite dimensional Hamiltonian systems and they are associated 
with infinite number of integrals of motion(see form example, Ablowitz \& 
Clarkson, [1991]).  One can associate with each of these soliton equations 
an infinite number of conservation laws.  For example for the KdV the 
first three reads
$$u_t+(3u^2+u_{xx})_x = 0,\eqno(50a)$$
$$(u^2)_t+(4u^3+2uu_{xx}-u_x^2)_x = 0,\eqno(50b)$$
$$(u^3-\frac{1}{2}u_x^2)_t+\left(\frac{9}{2}u^4+3u^2u_{xx}-6uu_x^2-
u_xu_{xxx}+\frac{1}{2}u_{xx}^2 \right )_x = 0.\eqno(50c)$$
The rest of them can be obtained recursively.

Furthermore, the existence of these infinite number of conservation laws can 
be associated with the existence of infinite number of generalized symmetries, 
namely the so called Lie-B\"{a}cklund symmetries, from which using Noether's 
theorem through suitable recursion operators the integrals can be obtained.  
For fuller details see for example, Bluman \& Kumei[1989].

One of the fundamental concepts which underlies the IST method of the solution 
is the interpretation that nonlinear evolution equations which are solvable 
by the IST scheme are completely integrable infinite dimensional Hamiltonian 
systems and IST can be thought of as a nonlinear transformation from physical 
variables to an infinite set of action-angle variables.  Such a description 
was developed for the KdV equation by Zakharov and Faddeev [1971] followed by 
others to different soliton equations (see Ablowitz \& Clarkson [1991]).

Considering the KdV equation, it can be written as 
$$u_t = \{u,H\}, \eqno(52)$$
where H is the Hamiltonian 
$$H = - \int_{-\infty}^{\infty} (u^3 -\frac{1}{2}u_x^2)dx, \eqno(53)$$
and the Poisson bracket between two functionals $A(\alpha)$ and $B(\beta)$ 
are defined by 
$$\{A(\alpha),B(\beta)\}  \equiv \int_{-\infty}^{\infty} \left \{ 
\frac{\delta A(\alpha)}{\delta u(x)} \; \frac{\partial}{\partial x} \;
\frac{\delta B(\beta)}{\delta u(x)} \right\}dx. \eqno(54)$$
Here  $\frac{\displaystyle \delta}{\displaystyle \delta u} $ stands for 
functional derivative.  Then defining the canonical coordinates 
$$ P_j = k_j^2, \; Q_j = -2\ln |C_j|, \; j=1,2\ldots,N, \eqno(55a)$$
$$ P(k) = k \pi^{-1} \ln|a(k)|^2, \; Q(k) = -\frac{i}{2}
\ln\left[\frac{b(k)}{\overline b(k)}\right], \; -\infty<k<\infty \eqno(55b)$$
where $k_j$ is the bound state eigenvalue of the associated Schr\"{o}dinger 
spectral problem, $C_j$ is the corresponding normalization constant and 
$a(k)$ and $b(k)$ are related to transmission and reflection coefficients, 
one finds 
$$ \{P_j, Q_j\} = \delta_{ij}, \; \{P_i,P_j\} = \{Q_i,Q_j\} = 0  \eqno(56)$$
$$ \{P(k), Q(k')\} = \delta (k-k'), \{P(k), P(k')\} = 
\{Q(k), Q(k')\} = 0. \eqno(57)$$
Then the Hamiltonian $H$ becomes
$$\hat H = -\frac{32}{5}\sum\nolimits_{j=1}^N P_j^{5/2}+
8 \int_{-\infty}^{\infty} k^3 P(k)dk,\eqno(58)$$
so that writing the equations of motion one finds that 
$$P(k,t) = P(k,0), \; Q(k,t) = Q(k,0)+8k^3 t, \eqno(59a)$$
$$P_j(t) = P_j(0), \; Q_j(t) = Q_j(0)-16P_j^{3/2}t. \eqno(59b)$$
Thus $P's$ and $Q's$ constitute an infinite set of action-angle variables and  
in this sense the KdV is a completely integrable Hamiltonian system.  Any other
soliton system can be shown to have exactly similar description.
\subsection { The Painlev\'{e} property}
In the earlier sections, we have seen that several interesting nonlinear 
dispersive wave equations admit soliton solutions so 
that the Cauchy initial value problem can be solved and they can be considered 
completely integrable (in the Liuoville sense). However the question arises
whether given a nonlinear partial differential equation, one can conclude
before hand that it is integrable and that linearization can be effected.
It is now well-recognized that a systematic approach to determine
whether a nonlinear pde is integrable or not is to investigate the
singularity structure of the solutions, in particular, their so-called
Painlev\'{e} property.  This approach, originally suggested by Weiss,
Tabor \& Carnevale [1983] (WTC), aims to determine the presence or absence
of movable noncharacteristic critical singular mainfolds
(of branching type, both algebraic and logarithmic, and essential singular 
type).  When the system is free from movable critical manifolds, the 
Painlev\'{e} property holds, suggesting its integrability.  Otherwise the 
system is nonintegrable in general.

The above development is a natural generalization of analyzing ordinary 
differential equations (odes) as per the movable critical singular points 
admitted by the solutions in the complex plane of the independent variable.  
Such a procedure was originally advocated by the mathematicians Fuchs, 
Painlev\'{e}, Gambier, Garnier, Chazy, Bureau and others  
and was applied successfully to the rigid body dynamics by S. Kovalavskaya
as mentioned in Sec.2.2 [Kruskal \& Clarkson, 1992].  A 
recent revival is due to the findings of Ablowitz, Ramani \& Segur [1980], 
who conjectured that similarity reductions of integrable soliton equations 
always lead to odes free from movable critical singular points [Lakshmanan \& 
Kaliappan, 1983].  In recent times several integrable dynamical systems have 
been identified to be free from movable critical points (see, for example 
Lakshmanan \& Sahadevan [1992]), thereby giving a useful criterion for 
integrability.
\subsubsection { Painlev\'{e} analysis}
Let us consider a NLEE of the form
$$u_t+K(u) = 0, \eqno(60)$$
where $K(u)$ is a nonlinear functional of $u(x_1,x_2,\ldots,x_M,t) = 
u({\bf x},t)$ 
and its derivatives up to order N, so that Eq.(60) is an Nth order nonlinear  
pde.  Then one may say that (60) possesses the Painlev\'{e} or P-property 
if the following conditions are satisfied.

(A) The solutions of (60) must be single-valued about the noncharacteristic 
movable singular mainfold.  More precisely, if the singular manifold is 
determined by
$$ \phi({\bf x},t) = 0, \; \phi_{x_i}({\bf x},t) \neq 0, \; \phi_t({\bf x},t) 
\neq 0,\; i=1,2,\ldots,M, \eqno(61)$$
and $u({\bf x},t)$ is a solution of the pde (60), then we have the Laurent expansion 
$$u({\bf x},t) = [\phi({\bf x},t)]^{-m} \sum\nolimits_{j=0}^{\infty} 
u_j({\bf x},t)\phi^j({\bf x},t), \eqno(62)$$
where $\phi({\bf x},t), \; u_j({\bf x},t)$ are analytic functions of 
$({\bf x},t)$ in a deleted 
neighborhood of the singular manifold (61), and m is an integer. 

(B) By the Cauchy-Kovalevskaya theorem the solution (62) should contain N 
arbitrary fucntions, one of them being the singular manifold $\phi$ itself 
and the others coming from the $u_j$'s.

Then the WTC procedure to test the given pde for its P-property essentially 
consists of the following three steps [Weiss et al., 1983]:

(i) Determination of leading-order behaviours,

(ii) Identification of powers j (resonances) at which the arbitrary fucntions 
can enter into the Laurent series expansion (62), and 

(iii) Verifying that at the resonance values $j$ a sufficient number of 
arbitrary fucntions exist without the introduction of movable critical 
singular manifold.

An important feature of the WTC formalism is that the generalized Laurent 
series expansion can not only reveal the singularity structure aspects of the 
solution and integrability nature of a given pde, but can also provide an 
effective algorithm which in most cases successfully captures all its 
properties, namely the linearization, symmetries and so on.

As a simple application, we illustrate the above aspects with KdV as an 
example.  Any other soliton system can also be likewise analyzed.   
For the KdV equation (17), we substitute the 
formal Laurent series expansion (62) around the singularity manifold 
$\phi(x,t) = 0$ and equate equal powers of $\phi$ to zero.  One finds 
that the exponent m=2 and that at $j=-1,4,6$ arbitrary functions can enter the 
power series (62).  Identifying the arbitrariness of $\phi$ with $j=-1$, 
recursively one finds
$$j = 0: u_0 = -2\phi_x^2,\eqno(63a)$$
$$j = 1: u_1 = 2\phi_{xx},\eqno(63b)$$
$$j = 2: \phi_x \phi_t+6u_2\phi_x^2+4\phi_x\phi_{xxx}
-3\phi_{xx}^2 = 0,\eqno(63c)$$
$$j = 3: \phi_{xt}+6u_2\phi_{xx}-2u_3\phi_x^2+\phi_{xxxx} = 0,\eqno(63d)$$
$$j = 4: \frac{\partial}{\partial x}(\phi_{xt}+6u_2\phi_{xx}
-2u_3\phi_x^2+\phi_{xxxx}) = 0.\eqno(63e)$$
Now it is clear that by the condition (63d), (63e) is always satisfied so that 
$u_4(x,t)$ is arbitrary.  Similarly one can derive the condition at $j=5$ and 
prove that at $j=6, \; u_6(x.t)$ is arbitrary.  The KdV equation is of third 
order, the Laurent series admits three arbitrary functions (without the 
introduction of a movable critical manifold) and so the Painlev\'{e} property 
is satisfied. 

Now, if the arbitrary fucntions $u_4$ and $u_6$ are chosen to be identically 
zero and, further, if we require $u_3 = 0$ then $u_j = 0, j \geq 3$, provided 
$u_2$ satisfies the KdV.  Thus we obtain the BT for the KdV in the form
$$u = (\log\phi)_{xx}+u_2,\eqno(64)$$
where $u$ and $u_2$ solve the KdV and $\phi$ satisfies (63c-d) with $u_3 = 0$.  
By a set of transformations, it is possible to show that the defining 
equations for $\phi$ are indeed equivalent to the linearizing equations (19),
as well as the bilinear equation (32).  Thus the analytic properties 
associated with the KdV equation can also be obtained from the Painlev\'{e} 
procedure as well.
The same procedure can be applied to any other NLEE to obtain its integrability
property in any dimension.

\subsection { Some applications of soliton concept}
The remarkable stability of soliton excitations in nonlinear dispersive 
systems caught immediately the imagination of physicists working in different 
areas of physics, apart from the traditional areas of hydrodynamics and fluid 
mechanics.  Innumerable applications of soliton concept in diverse areas of 
physics have been realized during the past three decades (see for example, 
Ablowitz and Clarkson [1991]; Lakshmanan [1995]).  We will consider 
here only a select few of them to illustrate the vast potentialities of the 
concept.  These include applications in magnetism, nonlinear optics, liquid 
crystals and elemenary particle physics. 
\subsubsection { Solitons in ferromagnets}
Spin excitations in ferromagnets are effectively expressed in tems of the 
Heisenberg's nearest neighbour spin-spin exchange interaction with additional 
anisotropy and external field dependent forces.  For the simplest isotropic 
case, the Hamiltonian for quasi-one dimensional ferromagnets is given by 
$$ H = -J  \sum\nolimits_{\{i,j\}}\overrightarrow {S_i}.\overrightarrow {S_j}  
\eqno(65)$$
where the spin operator $\overrightarrow {S_i} = (S_i^{\displaystyle x},
S_i^{\displaystyle y},S_i^{\displaystyle z})$ and $J$ is the   
exchange integral.  The Heisenberg equation of motion
$$\frac{d \overrightarrow {S_i}}{dt} = [S_i,H], \eqno(66)$$
in the long wavelength, low temperature (semiclassical $\hbar \rightarrow 0$) 
limit can be expressed in terms of classical unit vectors 
$$\frac{d \overrightarrow {S_i}}{dt} = \{S_i,H\}_{PB},\;{\overrightarrow {S_i}}^2 =1, 
\eqno(67)$$
where the Poisson brackets between two functions of spin can be defined as 
$$ \{A,B\}_{PB} = \sum\nolimits_i \epsilon_{\alpha\beta\gamma}\; \frac{\partial A}
{\partial S_i^\alpha} \; \frac{\partial B} 
{\partial S_i^\beta} \; S_i^\gamma.\eqno(68)$$
Correspondingly the equation of motion for the Hamiltonian (65) can be 
written as 
$$ \frac{d \overrightarrow {S_i}}{dt} = J\;\overrightarrow{S_i} \times
\{\overrightarrow S_{i+1}+\overrightarrow S_{i-1}\}. \eqno(69)$$
Additional interaction can also be included in the same way.  For example 
with a uniaxial anisotropy and external magnetic field along the z-direction, 
the Hamiltonian is 
$$ H = -J \sum\nolimits_{ \{i,j\} } \overrightarrow {S_i}.\overrightarrow
{S_j}+A \sum\nolimits_{i} \left(S_i^{\displaystyle z}\right)^2 -\mu \vec B.
 \sum\overrightarrow {S_i}, \eqno(70)$$
so that the equation of motion for the spins becomes
$$ \frac{d \overrightarrow {S_i}}{dt} = \overrightarrow{S_i} \times
\left\{ J\left(\overrightarrow S_{i+1}+\overrightarrow S_{i-1}\right)
-2AS_i^z \vec n +\mu \vec B\right\}\eqno(71)$$
where $\vec B = (0,0,B) $ and $\vec n=(0,0,1)$.

In the continuum limit, 
$$\overrightarrow {S_i}(t) \rightarrow \overrightarrow {S}(x,t), \;
\overrightarrow {S}_{i+1}(t) = \vec S(x,t)+a \frac{\partial \vec S}
{\partial x}+\frac{a^2}{2}\frac{\partial^2 \vec S}
{\partial x^2}+\cdots ,\eqno(72)$$ 
where a is the lattice parameter, the equation of motion (69) in the limit 
$a\rightarrow 0$ and after a suitable rescaling becomes
$$\frac{\partial \vec S}{\partial t} = \overrightarrow S \times \frac{\partial^2 
\overrightarrow S}{\partial x^2}. \eqno(73)$$
Similarly eq.(71) in the continuum limit becomes 
$$\frac{\partial \overrightarrow S}{\partial t} = \overrightarrow S \times 
(\frac{\partial^2 \overrightarrow S}{\partial x^2}+2AS^{\displaystyle z} 
\overrightarrow n + \mu \overrightarrow B). \eqno(74)$$

Spin equations of the type (73) or (74) are special cases of
the so called Landau-Lifshitz 
equation, which were derived originally by Landau and Lifshitz[1935] 
from phenomenological arguments.  It was not until 1977 that the 
complete integrable nature of many of these systems was realized.  In fact, 
by mapping eq.(73) on a moving space curve with curvature (Lakshmanan [1977])
$$ \kappa(x,t) = \left[\frac{\partial \vec S}{\partial x}. 
\frac{\partial \vec S}{\partial x}\right]^{1/2},\eqno(75)$$
and torsion
$$ \tau(x,t) = \kappa^{-2}\left(\vec S.\frac{\partial \vec S}{\partial x}
\times \frac{\partial ^2 \vec S}{\partial x^2}\right), \eqno(76)$$
which are respectively related to energy density and current density, eq.(73) 
gets mapped onto the ubiquitous nonlinear Schr\"{o}dinger equation (see 
Table I) (Lakshmanan [1977])
$$  iq_t+q_{xx}+2|q|^2q = 0, \eqno(77)$$
where the complex variable 
$$q(x,t)  = \frac{\kappa(x,t)}{2} \exp i\int_{-\infty}^x \tau dx'.\eqno(78)$$
Thus the spin equation (73) itself becomes a completely integrable soliton 
system with the one soliton solution for the energy density being given by 
$$ \left (\frac{\partial \overrightarrow S}{\partial x} \right )^2 =
\kappa^2(x,t) = 16 \eta^2 \;  sech (2 \eta (x-x_0)+8 \eta \xi t) \eqno(79)$$
and the spin component becomes 
$$ S^Z(x,t) = 1- \frac{2\eta^2}{(\xi^2+\eta^2)}\; sech^2 (2 \eta
[x-2 \xi t-\theta_0]). \eqno(80)$$
Also one can write down a Lax pair for (73) itself (Takhtajan, [1977])
$$L = iS \; and \; B = -iS\frac{\partial^2}{\partial x^2}-
iS_x\frac{\partial}{\partial x}, \; S=\overrightarrow S. 
\vec \sigma, \eqno (81)$$
\centerline {($\vec \sigma$: Pauli matrices)}

\noindent so that an IST analysis can be performed directly thereby again
proving the complete integrability of the spin system.  

Further the above analysis also shows that the spin system (73) is geometrically 
equivalent to the nonlinear Schr\"{o}dinger equation (Lakshmanan, [1977]) 
and that their eigenvalue problems are gauge equivalent (Zakharov and 
Takhtajan, 1979).  Since then the analysis of nonlinear excitations in 
magnetic system has been drawing considerable interest.  One finds that even 
in the presence of additional external magnetic fields or uniaxial or 
biaxial anisotropies the system continues to be completely integrable (for 
details see for example, recent reviews of Mikeska \& Steiner, [1991];
Kosevich etal, [1990]).
One can use the presence of such stable excitations to develop suitable 
statistical mechanics and obtain structure factors to compare with 
experimental results.  Effects of further interactions leading to 
integrability and nonintegrability and damping effects, etc., are some of the 
important topics of current interest which are being pursued vigorously.
\subsubsection { Solitons in optical fibers}
It is well known now that during the past 20 years or so optical fibers have 
revolutionized the world's telephone system.  By transforming speech into 
pulses of light and sending these pulses along ultra-clear glass fibres, 
communication engineers can pack thousands of telephone conversations into 
a single fiber. But in optical fiber communication, both dispersion and fiber
 loss are two important variables 
which decide the information capacity and distance of transmission.  An 
effective technology to overcome the problem of dispersion and fiber loss is 
being developed starting from the idea of soliton based communications 
first proposed by Hasegawa in 1973 [Hasegawa, 1989].  Solitons in optical 
fiber was first observed by Mollenauer [1990]) and the soliton laser was 
developed in 1984.  Since optical soliton arises as the balance between the 
nonlinear effect and the group velocity dispersion effect as in the case of 
any other soliton phenomenon, no distortion of the pulse takes place to a 
first order as a consequence of the dispersion.  As a result the optical 
soliton technology will be expected to make revolution in international 
communications in the next few years.  However, when the light intensity of 
the soliton decreases due to the fiber loss, the pulse width of the soliton 
transmission system also requires suitable reshaping of the pulse by 
suitable amplifiers.

Mathematically, optical soliton is the stationary solution of the initial 
boundary value problem of the nonlinear Schr\"{o}dinger equation for the 
light intensity $E(z,t)$
$$ iE_z+i\kappa \kappa' E_t -\frac{\kappa''}{2}E_{tt} +
\frac{\omega n_2}{c}|E|^2E = 0, \eqno(82)$$
where $\kappa '(=\frac{\displaystyle 1}{\displaystyle v_g})$ is the inverse 
of the group velocity, 
$\kappa '' = \frac{\displaystyle \partial^2 \displaystyle \kappa}
{\displaystyle \partial^2 \displaystyle \omega^2}$ is the group velocity
dispersion coefficient and $n_2$ is the nonlinear refraction coefficient,
c is the velocity of light and $\omega$ is the carrier frequency.
Under the transformation $\tau = t-\frac{\displaystyle z} {\displaystyle v_g}$
eq.(82) can be written as
$$ iE_z -\frac{\kappa''}{2}E_{\tau \tau} + \frac{\omega n_2}{c} |E|^2E = 0.
\eqno(83)$$
The one soliton solution then can be given as
$$ E(z,t)=\sqrt{\frac{c^2}{\omega n_2}}.\eta.sech \eta (\tau+\kappa z-
\theta_0).\exp \left \{ -i\kappa \tau+\frac{i}{2}(\eta^2-\kappa^2)z \right \},
\eqno(84)$$
where $\eta$, $\theta_0$ and $\kappa$ are constant parameters. The actual
derivation of eq.(83) follows by starting from Maxwell's equations for
the propagation of intense electromagnetic radiation along the optical fibre,
which is a silica dielectric medium. Using appropriate slowly varying envelope
approximations for propagation along the fibre direction and taking into
account the necessary nonlinearity for the refractive index, eq.(83) can be
obtained. (For details, see for example Agarwal [1995]).

\subsubsection{Solitons in Liquid Crystals}
Liquid crystal is a state of matter intermediate between liquid and
crystal. The material is optically anisotropic and can
flow atleast in one spatial dimension. The molecules of the organic compound
showing liquid crystal phases can be either rod-like, disc-like or bowl
like in shape. At low temperature, both the orientations and positions of
the molecules are in order (long-range) and so we have the crystal phase.
At high temperature, both types of degrees of freedom are in disorder
and the material is in the isotropic liquid phase. However, within a certain
temperature range, there may exist an intermediate phase (mesophase) in which
the orientations are in order, but the positions are in disorder. Such an
intermediate phase (mesophase) is called ``nematic" and the other known
mesophases are smectic, cholosteric and ferroelectric.

In liquid crystals, since the molecules have both orientational and
translational degrees of freedom, the hydrodynamic equations of motion are
coupled nonlinear equations on $\vec n$ and $\vec v$ where `$\vec n$' is the
director, a unit vector representing the average orientation of the
molecules and $\vec v$ is the velocity of the centre of mass of the
molecules and both $\vec n$ and $\vec v$ are functions of space and time.
The orientation of the molecules can be detected optically as it is localized
and orientational waves can be observed easily by the naked eye. This
provides a convenient means of measuring the wave. Thus, the identification
of solitons in liquid crystals play an important role in the switching
mechanism of some ferroelectric crystal displays. (For details see for
example, Lam and Prost [1991]; Lam [1995]).

 (i) Similar to domain (Bloch) wave in a ferromagnet, a soliton in a liquid
crystal is a smooth, localized state linking up two uniform states at the
two far ends.

 (ii) For propagating a soliton in a liquid crystal, the damping of the
director is heavy resulting in the overdamped case in the equation of
motion for the director. Thus, liquid crystals are generally nonintegrable
systems and the solitons involved are usually just solitary waves.

 (iii) In liquid crystals, reorientation of the molecules can
generally induce fluid flows.

Considering an one dimensional shear of nematic, the orientation of the
molecule $\theta$ obeys the overdamped sine-Gordon equation
$$k\theta_{xx}-r_1\theta_t+\frac{s}{2}(r_1-r_2\cos2\theta)=0. \eqno (85)$$
(Here $s=\frac{dv}{dx}$ is the shear and $r_1$, $r_2$ are viscosities, $k$ is
the elastic constant). The above equation has two uniform steady states, $\pm
\theta_0$, $\theta_0=\frac{1}{2}\cos^{-1} \frac{r_1}{r_2}$. It is then possible to
have a localized configuration of the molecules. When this configuration
travels without distortion, we have a soliton.

As liquid crystals are also nonlinear optical materials (dielectrics), optical
solitons can also be found in liquid crystals. When liquid crystals are
subjected to an electric field E perpendicular to $\vec n$, they tend to align
parallel to the field only if the electric field is greater than a threshold
value $(E>E_{th})$ and the corresponding equation of motion is given by
$$A_{yy}+a_2\omega^2|A|^2A+2i\omega a_1A_t-2ikA_z=0, \eqno (86) $$
where E is assumed to be linearly polarized along $x$ direction, propagating
along $z$ so that $E=A(y,z,t)e^{i(\omega t-kz)}$. When $A_t=0$, the above equation
reduces to NLS equation (self-focussing case). When $A_z=0$ which
corresponds to the case of self-modulation, equation (86) is scaled with a new 
time variable to NLS equation. In both the cases optical solitons exist.

\subsubsection{Solitons in particle physics}
The various models to describe different types of elementary particles and 
their interactions are generically nonlinear field models at a classical
level (recall Einstein's view quoted in Sec. 2.5 in this regard), which
then needs to be second quantized. In the (1+1) dimensional cases they
often reduce to the well known integrable soliton systems such as the 
sine-Gordon equation, $\sigma$-model equation, Lund-Regge equation and
so on. One may cite many examples, where nonlinearity arises (Note that
we have already discussed the Skyrme model briefly in Sec.3), which are
only illustrative and not exhaustive.

1.\underline {Higgs mechanism:} In the so called standard model of electroweak
gauge theory in order to give masses for the (weak) gauge bosons, a complex
field with Lagrangian $L=-\frac{1}{2}\partial_{\mu}\phi \partial^{\mu} \phi +
\mu^2 \phi^2+\lambda \phi^4 $ is introduced. Looking at the vacuum solution 
$<\phi>=\sqrt{-\frac{\mu}{\lambda}}$ and expanding $\phi \longrightarrow <\phi>
+\phi $, the fluctuation $\phi$ becomes the Higgs field. A search for the Higgs
particle is still elusive.

2.\underline {Non-abelian theory:} It is a manifestly nonlinear theory. The
$SU(2)$ gauge fields $A_{\mu}^a \; (a=1,2,3)$ or $SU(3)$ gluons $A_{\mu}^a \; 
(a=1,\cdots,8)$ have cubic and quartic interactions. The field strength is 
$$ F_{\mu \nu}^a = \partial_{\mu}A_{\nu}^a-\partial_{\nu}A_{\mu}^a+f^{abc}
A_{\mu}^bA_{\nu}^c\eqno (87)$$
where $f$'s are the structure constants of the gauge group. Even 
in the absence of fermions, a free Yang-Mills theory is difficult to solve. 
In this case the Lagrangian is 
$$L=-\frac{1}{4}F_{\mu\nu}^aF^{\mu\nu a} \eqno (88)$$
and the equation of motion is
$$D_{\mu}^{ab}F_{\mu\nu}^b =0, \; D_{\mu}^{ab}=\partial_{\mu}\delta^{ab}
+f^{abc}A_{\mu}^c. \eqno (89)$$
An interesting special case is the self-dual Yang-Mills equation where 
$$F_{\mu\nu}=\frac{\partial A_{\nu}}{\partial x_{\mu}} -\frac{\partial 
A_{\mu}}{\partial x_{\nu}}-\left [ A_{\mu},A_{\nu} \right ]. \eqno (90) $$
There has been considerable work on the integrability of this equation and
it is conjectured that all or atleast most of the soliton eqautions are special 
reductions of it [see for example, Ward, 1985; 1986]. These four dimensional
equations arise in the study of field theory and relativity. The SDYM equation
is regarded as being ``integrable" as a consequence of the so called 
``twistor" representation relating their solutions to certain holomorphic 
vector bundles.

3.\underline {Gravitation theory:} Of course the gravitational field
equations even in the absence of matter $(T_{\mu\nu}=0)$
$$ R_{\mu\nu}-\frac{1}{2}g_{\mu\nu}R=0 \eqno (91) $$
is highly nonlinear. Classical solutions such as Swarzchild, Kerr, etc. are
well known. Recently Einstein's field equations with axial symmetry in the
form of Ernst equation was found to be integrable. There are other solitons
bearing cases also known in the literature.

4.\underline {String theory:} Much work has been carried out on string theory
for quantum chromodynamics. These are highly complex nonlinear field 
equations and sophisticated mathematical results have come out of these
studies. However we do not consider any details here. Interested reader
may refer, for example, [Witten, 1985].

\subsection { Solitons in higher dimensions}
Do solitons or their localized counterparts exist in higher spatial dimensions 
too?  If so, what are their characteristic features and when do they occur 
and what are their ramifications?  These are of paramount physical importance 
as most natural systems are higher spatial dimensional in nature.  Following 
the original works of Zakharov \& Manakov [1985] and the references therein, 
Ablowitz, Fokas \& coworkers [1983] and references therein, there has been 
intense activity in understanding nonlinear dispersive wave equations in 
higher dimensions during the past ten years or so.  The Kadomtsev-
Petviashvile(K-P), Davey-Stewartson(D-S) and Ishimori equations are some of 
the well studied (2+1) dimensional systems which are interesting
generalizations
of the (1+1) dimensional KdV, nonlinear Schr\"{o}dinger and Heisenberg 
ferromagnetic spin (vide Table II) equations.  Depending on the sign of the 
coefficients of certain terms, these equations are also further classified 
as KPI, KPII, DSI and DSII, etc., Naturally, these (2+1) dimensional equations 
are richer in structures, where boundary conditions play a crucial role
[Ablowitz \& Clarkson, [1991]).  Some of the (2+1) dimensional nonlinear 
coherent structures which have been invented in recent times are 

 (i) \underline {line solitons} (for example KP and DS) which do not decay 
in all directions, but there exists certain lines on which the solutions 
are bounded but nondecaying.

 (ii)\underline {lump solitons} (for example KPI and DSII) which are localized 
but decay algebraically and do not in general suffer a phase-shift under 
collision.

 (iii)\underline {dromions} (for example DSI) which are driven by boundary 
effects, being localized and exponentially decaying excitations which in 
general undergo amplitude and velocity changes under collision but whose 
total number and energy are conserved [Fokas \& Santini, 1990].  The  
explicit forms of some of these excitations are also given in Table II and 
some of them are displayed in Figs. 10-12.

We have noted above that certain nonlinear evolution equations in
higher spatial dimensions are also linearizable.  Thus they can also be 
analyzed through the three-step procedure indicated earlier.  However, 
scattering analysis in higher dimensions is much more complicated than that 
in one dimension, and a new approach is required to deal with them.  The 
$d$-bar approach [Ablowitz \& Fokas, 1983; Beals \& Coifman, 1989; 
Zakharov \& Manakov, 1985] treats the scattering problem in any dimensions as 
a $d$-bar problem of analytic functions in complex variable theory.  With 
this new interpretation, it is possible to approach the inverse scattering 
analysis for evolutions both in (1+1)- and (2+1)-dimensions in a unified way.  
\subsubsection { The $d$-bar problem}
Given an analytic function 
$$ f(z) = u(x,y)+iv(x,y),\\
z=x+iy,\; u,v,x,y \in {\cal R}, \eqno(92)$$
the Cauchy-Riemann conditions
$$\frac{\partial u}{\partial x} = \frac{\partial v} {\partial y}, \;
\frac{\partial v}{\partial x} = -\frac{\partial u}{\partial y}\eqno(93)$$
can be recast in the so-called $d$-bar form
$$\frac{\partial f}{\partial \bar {z}} = \bar{\partial}f = 0, \eqno(94)$$
where $\partial/ \partial \bar {z} = \bar{\partial}z = \bar \partial =
(1/2)[(\partial/ \partial x)+i(\partial/ \partial y)]$.  Thus $\bar{\partial}f 
= g(\neq 0)$ is a measure of the nonanalyticity of the function $f$.(Example: 
$f(z) = \bar{z}$).

Then the $d$-bar  problem is whether, given the $d$-bar data
$\bar{\partial}f$, one can invert it to get the function $f(z)$.  The answer 
is, in general, yes.  The procedure is to make use of the generalized 
Cauchy integral formula
$$f(z) = \frac{1}{\pi}\int_{-\infty}^{\infty}\int_{-\infty}^{\infty} 
\frac{1}{(z-\zeta)}\frac{\partial f}{\partial \bar{\zeta}}d\zeta_R d\zeta_I+
\frac{1}{2\pi i}\int_C \frac{f(\zeta)}{(z-\zeta)}d\zeta,\eqno(95)$$
where the last term is typically the identity if $f$ is normalized to unity 
for large z.

\subsubsection { General set up of IST method}
Considering the linear eigenvalue problems associated with the NLEEs in (2+1) 
dimensions, they may be written as 
$$P(\lambda)\psi({\bf x},\lambda) = Q({\bf x})\psi({\bf x},\lambda), \; 
{\bf x} = (x,y), \eqno(96)$$
where $P(\lambda)$ is a matrix-valued linear differential operator in $x,y 
\in {\cal R}^2$ and $analytic$ in $\lambda \in {\cal C}$.  Here $Q({\bf x})$ 
is the ``potential'' and we assume $Q({\bf x}) \rightarrow 0$ as  
$|\bf x| \rightarrow \infty$.  The (1+1) dimensional case is obviously a special 
case of (96).  For example 
$$P(\lambda) = \frac{d^2}{dx^2}+\lambda ^2, \eqno(97)$$
for K-dV equation, and 
$$P(\lambda) = \frac{\partial ^2}{\partial x^2}+i\frac{\partial}{\partial y},
\eqno(98)$$
for KPI.

Since the potential $Q \rightarrow 0$ as $|\bf x| \rightarrow \infty$, we can 
easily find the asymptotic form $\psi({\bf x},\lambda) \rightarrow 
e_{\lambda}({\bf x})$ so that 
$$P(\lambda)e_{\lambda}({\bf x}) = 0, \; |\bf x|\rightarrow \infty .
\eqno(99)$$
Then letting 
$$\psi({\bf x},\lambda) = m({\bf x},\lambda)e_{\lambda}({\bf x}),
\eqno(100)$$
so that 
$$ m({\bf x},\lambda) \rightarrow I \; as |{\bf x}|\rightarrow \infty, 
\eqno(101)$$
we have the modified eigenvalue problem
$${\cal P}_{\lambda}m({\bf x},\lambda) = Q({\bf x})m({\bf x},\lambda). 
\eqno(102)$$
For example, for K-dV this reads 
$$ m_{xx}+2i\lambda m_x = -um, \eqno(103)$$
and for KP-I we have 
$$m_{xx}+im_y+2i\lambda m_x = -um.\eqno(104)$$
\subsubsection { Direct scattering}
Considering the Green's function associated with the eigenvalue problem 
(96),
$$ {\cal P}_{\lambda}G_{\lambda}({\bf x}-{\bf x'}) = \delta({\bf x}-
{\bf x'}),\eqno(105)$$
subject to the boundary condition (101), we have formally
$$m({\bf x},\lambda) = I+\int_{\Omega} G_{\lambda}({\bf x}-{\bf x'})
Q({\bf x'})m({\bf x'},\lambda)d{\bf x'}\\
=I+G_{\lambda}*(Qm), (*:convolution)\eqno(106)$$
where $I$ is the unit matrix and $\Omega$ is the domain of integration.
Rewriting (106), we have
$$(I-G_{\lambda}*Q)m = I.\eqno(107)$$
Assuming now that formally the inverse of $(I-G_{\lambda}*Q)$ exists 
(which requires control over $G$ and the potential $Q$), we can write the 
formal expression for the eigenfucntion,
$$ m({\bf x},\lambda) = (I-G_{\lambda}*Q)^{-1}.I.\eqno(108)$$
Considering the analytic behaviour of the eigenfunction in the complex 
$\lambda$ plane, its nonanalyticity is given by the $d$-bar data
$$\bar {\partial}m = \bar {\partial}[I-G_{\lambda}*Q]^{-1}. I =
(I- G_{\lambda}*Q)^{-1}(\bar {\partial}G_{\lambda})*(Qm).\eqno(109)$$
 On the other hand, we have from Eq.(102)
$${\cal P}_{\lambda}.\bar {\partial}m = Q.\bar{\partial}m,\eqno(110)$$
using the analyticity property of ${\cal P}_{\lambda}$.  Thus 
$m({\bf x},\lambda)$ and $\bar {\partial}m$ are simultaneous eignefunctions 
of ${\cal P}_{\lambda}$.  From the completeness property of $m$, we have 
$$\bar{\partial}m = Tm,\eqno(111)$$
where T is the scattering operator obtained through a superposition of 
eignfunctions.  Comparing (109) and (111), T can be expressed as
$$T = [I-G_{\lambda}*Q]^{-1}(\bar{\partial}G_{\lambda})*Q.\eqno(112)$$

In typical one-dimensional problems such as the K-dV, sine-Gordon, etc. 
equations associated with the Z-S and AKNS eigenvalue problems, Eq.(112) turns 
out to be a local equation in $\lambda$ and $m({\bf x},\lambda)$ is 
sectionally holomorphic ( in the upper and lower half $\lambda$ planes with 
finite number of isolated singular points), so that we have a local Riemann-
Hilbert problem with singularities.  Consequently T turns out to be the 
scattering matrix in these cases [Ablowitz \& Fokas, 1983].

In the case of two-dimensional systems such as the KPI, DSI, Ishimori I 
equations, (112) turns out to be nonlocal even though $m$ is sectionally 
holomorphic and as a result we have a nonlocal Riemann-Hilbert problem [
Ablowitz \& Clarkson, 1991].

On the other hand, in problems like KP II and DSII, $m({\bf x},\lambda)$ is 
nowhere analytic and so one has to deal with a full $d$-bar problem in these 
cases[Ablowitz \& Clarkson, 1991].
\subsubsection { Time evolution}
Given the ``potential'' matrix $Q({\bf x},0)$ in (96), which is available 
from the initial data, Eq.(112) defines the scattering data.  In order to 
find the evolution of the scattering data, one essentially uses the time 
evolution of the eigenfunction of the form $\psi _t = B\psi$, where 
$B$ is a matrix linear differential operator in which the unknown 
$Q({\bf x},t)$ occurs as coefficients.

For asymptotically vanishing potentials, $Q\rightarrow 0$ as $|\bf x|\rightarrow
\infty,\; B\rightarrow B_0$, which is independent of $Q$, as a result $T$ 
evolves in a simple way, satisfying linear orindary/partial differential 
equations.  Consequently, the scattering data operator can always be 
determined in principle at an arbitrary time without any difficulty in all 
the cases where the potential vanishes at $\infty$, once they are known at 
$t=0$.

\subsubsection { Inverse scattering}
Given the scattering data operator T and the nonanlyticity of the function 
$\bar {\partial}m$, one can carry out an inverse scattering analysis[Beals \& 
Coifman, 1989] to retrive the function $m$ by invoking the generaized Cauchy 
integral formula (95), from which $Q$ itself can be obtained uniquely.

Formally this can be done as follows.  Let $C$ be the inverse of 
$\bar {\partial}m$.  Then taking into account the normalization of m as 
$\lambda \rightarrow \infty$, we can write 
$$m({\bf x},t,\lambda) = I+C.\bar{\partial}m $$
 $$ = I+C.Tm $$
 $$ = I+\frac{1}{\pi}\int_{-\infty}^{\infty}\int_{-\infty}^{\infty}\frac{1}
{(\lambda-\zeta)} \times \left (\frac{\partial m}{\partial \bar\zeta}\right )
d\zeta_R d\zeta_I $$
 $$ = I +\frac{1}{\pi}\int_{-\infty}^{\infty}\int_{-\infty}^{\infty}\frac{1}
{(\lambda-\zeta)} \times (Tm)d\zeta_R d\zeta_I.\eqno(113)$$
When T is known, Eq.(113) gives the eigenfunction $m({\bf x},t,\lambda)$.  
Finally, to obtain the potential Q({\bf x},t), one looks for the asymptotic 
expansion of $m({\bf x},t,\lambda)$ for large $\lambda$ in the form 
$$m({\bf x},t,\lambda) = I + \frac{m_1({\bf x},t)}{\lambda}+0\left(\frac{1}
{\lambda ^2} \right). \eqno (114)$$
When this is used in the eigenvalue problem (96), one can express the potential 
Q in terms of the coefficient $m_1({\bf x},t)$.  On the other hand, using 
(114) in (113), $m_1$ and hence the potential can be expressed in terms of the 
scattering data.  Thus the inverse problem can be solved uniquely, thereby 
solving the Cauchy initial value problem in both (1+1)- and (2+1)-dimensions.  
Some of the simplest solutions obtained are given in Table II for KP and 
DS equations.

Finally what about extension of IST to (3+1) dimensions? There seems to
be serious difficulties in the inverse procedure due to certain
constraints on the scattering data. The problem remains open at present.

\subsection{ Bifurcation and chaos in physical systems}
Integrable nonlinear systems like the soliton systems described above, or
other finite degrees of freedom systems (for details see for example,
Lakshmanan \& Sahadevan [1993]) or diffusive systems such as the Burger's
equation, are all though very important from a physical point of view still
relatively rare and almost measure zero in number compared to the totality
of nonlinear systems.  Often they are said to be exceptions rather than rule,
in spite of the invention of a large number of them along with their nice
properties.  Under perturbations most of these integrable systems become
non-integrable.  For finite degrees of freedom Hamiltonian systems, often
KAM theroem becomes relevant under such circumstances(see for example
Lichtenberg \& Lieberman, [1983]).  However most systems behave in a much
more intricate way when the nonlinearity is increased or the KAM theorem is
violated.  We have already seen earlier in sec. 3.5 that Hamiltonian systems
such as the Henon-Heiles model can show sensitive dependence on initial
conditions depending upon the strength of nonlinearity, exhibiting chaotic
motions.  Similarly, systems like Lorenz system (sec. 3.6) show dissipative
chaos(Lichtenberg \& Lieberman [1983]).  During the past two decades, an
explosion of research has firmly led to the acceptance of chaos as an
ubiquitous and robust nonlinear phenomenon frequently encountered in nature
(Drazin [1992], McCauley[1993], Mullin[1993]) and the concept has permeated
almost all branches of science and technology.  The field is growing into a
stage where the initial surprises associated with the phenomenon are waning
and new understandings are appearing, while actual controlling and harnessing
of chaos are being contemplated.

The net result of the investigations on chaotic nonlinear dynamical systems
since 1970 is that the notion of complete predictability has given way to
deterministic chaos or randomness for suitable nonlinear dynamical systems.
Given an N-particle system with masses $m_i (i=1,2\ldots,N)$ and (constraint
free) forces $\vec {F}_i$ acting on them, the
state of the system is in general expected
to be uniquely specified by solving the set of second order ordinary
differential equations
$$ m_i \frac{d^2 \vec {r}_i}{dt^2} = \vec {F}_i (t,\vec {r}_1,\vec {r}_2,\ldots,
\vec {r}_N,\frac{d \vec {r}_1}{dt}\ldots,\frac{d \vec {r}_N}{dt}),\;
i=1,2,\ldots,N, \eqno(115)$$
for prescribed 6N initial conditions $\vec {r}_i(0)$ and $\frac
{\displaystyle d \displaystyle \vec {r}_{\displaystyle i}}
{\displaystyle {dt}}\Big |_{\displaystyle {t=0}}$.  This Laplace dictum that
for ``a superintelligence nothing could be
uncertain and the future as the past, would be present to its eyes"[Gleick,
1987] is already flawed: (i) when $N$ is large, one requires a
statistical description and (ii) when quantum effects are present 
uncertainities can arise even in the simultaneous prescription of
initial conditions necessitating a quantum description.  Now the
further advance in our understanding of the dynamics is that even when the
above two limitations are absent, depending on the nature of the forces in
eq.(70), that is whether $\vec{F}_i $ is linear or nonlinear, new
uncertainities can arise leading to deterministic randomness or chaos.  For
appropriately chosen nonlinear force $\vec{F}_i $ the system can show
sensitive dependence on initial conditions, which can never happen when
$\vec{F}_i $ is linear, leading to 
exponential divergence of nearby trajectories, a possibility already foreseen
by Poincar\'{e} (Sec.2.3) during the beginning of the century.  The physical
consequence is the butterfly effect of Lorenz(sec 3.6).

Since in this meeting, the various individual aspects of chaos are well
discussed, we include here only a very brief account of them for completeness.
For details the readers are referred to the various other articles in this
book.

\subsubsection {Chaos in dissipative and conservative nonlinear systems}
We have already seen in sec.3 that the Henon-Heiles and Lorenz systems are
prototype of Hamiltonian and dissipative chaos.  Since then much
understandings have been achieved on both types of chaos.

{\bf {(i) Dissipative systems}}:
The time evolution of these systems contracts volume  in  
phase-space(the   abstract   space   of   state   variables)   and 
consequently trajectories approach asymptotically either a chaotic 
or a non-chaotic attractor.  The latter may be a fixed  point,  a 
periodic limit cycle or a quasiperiodic attractor.  These and  the 
chaotic attractors are  bounded  regions  of  phase-space  towards 
which the trajectory  of  the  system,  represented  as  a  curve, 
converges in the course of long time evolution.  Bifurcation 
or qualitative changes of periodic attractors can occur leading to 
more complicated and chaotic structures as a control parameter  is 
varied.  Different routes to the onset of chaos have been identified(see
for example, Lichtenberg \& Lieberman [1983], Drazin [1992]).

          The chaotic attractor is typically neither a point nor a 
curve but  a  geometrical  structure  having  a  self-similar  and 
fractal (often multifractal) nature.  Such chaotic attractors  are 
called  strange  attractors.   Many  physically  and  biologically 
important nonlinear dissipative systems, both in  low  and  higher 
dimensions,  exhibit  strange  attractors  and  chaotic   motions.  
Typical examples are  the  various  damped  and  driven  nonlinear 
oscillators, the Lorenz  system,  the  Brusselator  model 
, the Bonhoeffer- van der Pol oscillator, the piecewise 
linear electronic circuits and so on (see for example, Lakshmanan \&
Murali [1996]).

{\bf {(ii) Conservative or Hamiltonian systems}}:
          Nonlinear systems of conservative  or  Hamiltonian  type 
also exhibit often chaotic  motions.  But  here  the  phase 
space  volume  is  conserved  and  so  no  strange  attractor   is 
exhibited.  Instead, chaotic orbits tend to visit all parts  of  a 
subspace  of  the  phase-space  uniformly.   The  dynamics  of   a 
nonintegrable conservative system is  typically  neither  entirely 
regular nor entirely irregular, but the phase-space consists of  a 
complicated mixture of regular and irregular components.   In  the 
regular region the motion is quasiperiodic and the orbits  lie  on 
tori while in the irregular  regions  the  motion  appears  to  be 
chaotic but they are not attractive in nature.   Typical  examples 
include  coupled  nonlinear  oscillators,   Henon-Heiles   system, 
anisotropic Kepler problem and  so  on (see for example Drazin [1992],
McCauley [1993]).

\subsubsection { Quantum chaos}
The deterministic randomness or chaos exhibited by generic nonlinear
dynamical systems has been found to present significant practical and
philosophical implications, and probably limitations as well, in the
description of microscopic world.  There is no doubt that quantum theory
is a more accurate description of nature.  However, Bohr's correspondence
principle requires that in the appropriate limit the remnance of signatures
of (classical) chaos (of macroscopic world), namely the exponential divergence
of nearby trajectories and the intrinsic uncertainity due to nonlinearity,
should follow, barring unforeseen singularities in the $\hbar \rightarrow 0$
limit ($h$:Planck's constant), which might prevent the smooth transition
from quantum mechanics to classical mechanics.  Search for such quantum
manifestations of classical chaos in the practical sense, which goes by the
terminology `quantum chaos' or `quantum chaology', has recently attracted
considerable interest (for details see for intstance, Gutzwiller[1990],
Nakamura[1993], Reichl[1992]).

For example, one might look for possible fingerprints of chaos in the
eigenvalue spectrum, wavefunction patterns and so on.  In particular, by
looking at the short range correlations between energy level (spacings) of a
large class of quantal systems such as billiards of various types, coupled
anharmonic oscillators, atomic and molecular systems, it has been realized
that there exists generically a universality in the spacing distribution of the
quantum version of classically integrable as well as chaotic systems.  For
regular systems nearest-neighbour spacings follow a Poisson distribution,
while chaotic systems follow either one of the three universality classes,
depending on the symmetry and spin.  These universality classes correspond
to Gaussian Orthogonal Ensemble (GOE) or Wigner statistics, Gaussian
Unitary Ensemble (GUE) statistics or Gaussian Symplectic Ensemble (GSE)
statistics, similar to ones which occur in random matrix theory of nuclear
physics.  For near integrable and intermediate cases the level distributions
are found to satisfy Brody or Berry-Robnik or Izrailev statistics(Reichl[1992]).

In recent times it has been found that highly excited Rydberg atoms and
molecules(which are effectively one-electron systems) under various
external fields are veritable goldmines for exploring the quantum
aspects of chaos.  These systems are particularly appealing as they are
not merely mathematical models but important physical systems which can
be realized in the laboratory.  Particular examples are the hydrogen
atom in external magnetic fields, crossed electric and magnetic fields,
van der Waals force, periodic microwave radiation and so on.  These
studies seem to have much relevance to the understanding of the so
called mesoescopic systems and such investigations are of high current
interest in nonlinear physics.

\subsubsection { Controlling, synchronization and secure communication}
The above studies make it clear that chaos is ubiquitous in nature and
that it is intrinsically unpredictable and sensitively dependent on
initial conditions so that nearby trajetories diverge exponentially. 
Consequently the phase trajectories (in the phase space) can take
complicated geometrical structures, for example a fractal structure for
typical dissipative chaotic system.  Naturally one would expect such a
complex motion cannot be controlled or altered by minimal efforts
unless drastic changes are made to the structure of the system. 
Surprisingly, recent investigations in this direction have clearly
demonstrated (see for example, Shinbrot etal [1993], Lakshmanan \&
Murali [1996]) that not only can chaotic systems be tamed or controlled
by minimal preassigned perturbations to avoid any harmful effects to the
physical system under consideration but controlling can be effected in a
purposeful way to make the system evolve towards a goal dynamics. 
Numerous control algorithms have been devised in recent times, many of
which have been experimentally verified, for controlling chaos and they
broadly fall under two categories, (i) feedback and (ii) nonfeedback
methods, which effectively use the fact that the chaotic attractor
contains infinite number of unstable periodic orbits which can then be
chosen suitably and controlled for regular motion.

Another but related consequence of sensitive dependence on initial
conditions is that two identical but independently evolving chaotic
systems can never synchronize to be in phase and in amplitude, as any
infinitesimal deviation in the starting conditions (or the system
specification) can lead to exponentially diverging trajectories making
synchronization impossible.  Contrast this with the case of linear
systems (and also regular motions of nonlinear systems), where the
evolution of two identical systems can be very naturally synchronized. 
In this connection, the recent suggestion of Pecora and Carroll[1993]
that it is possible to synchronize even chaotic systems by introducing
appropriate coupling between them has changed our outlook on chaotic
systems, synchronization and controlling of chaos, paving ways for new
and exciting technological applications: spread spectrum communications
of analog and digital signals(For details see for example, Lakshmanan \&
Murali, [1996]).

\section{Outstanding problems and future outlook}
In the earlier sections, we have traced briefly the salient feaures of the
development of various topics in nonlinear physics and their ramifications,
ultimately leading to the twin concepts of solitonic and chaotic structures.
While the soliton excitations are predominant in one space and one time
dimensional systems, the chaos phenomenon is well studied (at least
numerically) for low degrees of freedom systems. It is obvious that one has
touched only the tip of the iceberg as far as nonlinear systems are concerned
and our present understanding is confined to a narrow range of them. The
nature of excitations in physically relevant higher spatial dimensional
systems, the transition from integrable regular systems to nonintegrable 
and chaotic systems and the formation of spatio-temporal patterns on
perturbations of soliton systems are some of the important poblems to be
tackled in the next few decades. Also the definition of integrability,
particularly in the complex plane and its relation to real time
dynamical behaviour, is one of the most import fundamental notions to be
understood and extended to nonintegrable and chaotic situations. Many
new technologies which are in the process of unfolding as a result of
the various applications of the notions of solitons and chaos remain to
be harnessed to their full potentialities, in such areas as
magnetoelectronics, information technology, secure communications and
so on. In this section, we will focus briefly on some of these topics
with a view to point out the problems and potentialities of nonlinear
physics in the next few decades.
\hskip 40pt
\subsection {Integrability and chaos}
We pointed out in the previous section that soliton equations may be
considered as completely integrable infinte dimensional Hamiltonian systems.
From another point of view we also saw that solutions are meromorphic and
free from movable critical singular manifolds (Painlev\'{e} property). From
yet another point of view, the existence and uniqueness of their solutions
can be established. It is not only the Hamiltonian type soliton equations
which are known to be integrable. The nonlinear diffusive Burger's equation
$$u_t+uu_x=\gamma u_{xx} \eqno (116) $$
is linearizable in the sense that the celebrated Cole-Hopf transformation
(see for example, Sachdev [1987])
$$ u=-2\gamma \frac {v_x}{v} \eqno (117) $$
converts eq.(116) into the linear heat equation
$$ v_t+\gamma v_{xx}=0 \eqno (118) $$
and so may be considered to be integrable. Similarly the nonlinear diffusive
equation
$$ u_t+u^2 u_x + Du^2 u_{xx}=0   \eqno (119) $$
is known to be linearizable and possesses infinite number of Lie-B\"{a}cklund
symmetries (Fokas \& Yorstos [1982] ). Eqs.$(116)$ and $(119)$ also satisfy 
the Painlev\'{e}
property. Similarly for finite degrees of freedom, there exists both integrable
Hamiltonian and dissipative systems (Lakshmanan \& Sahadevan [1993]; Ramani,
Grammaticos \& Bountis [1989]), obeying the Painlev\'{e} property.

\noindent \underline {Examples}:

\noindent 1) Two coupled anharmonic oscillators:

$$\ddot x+2(A  +2\alpha x^2+\delta y^2)x=0, \eqno (120a) $$
$$\ddot y+2(B  +2\beta y^2+\delta x^2)y=0. \eqno (120b) $$

Integrable cases:

i)$A=B$, $\alpha=\beta$, $\delta=6 \alpha$

ii)$\alpha=\beta$, $\delta=2 \alpha$

iii)$A=4B$, $\alpha=16 \beta$, $\delta=12 \beta$

iv)$A=4B$, $\alpha=8\beta$, $\delta=6 \beta$

\noindent 2) Lorenz system: Eq.(29)

\noindent Integrable case:

i)$\sigma=\frac{1}{2}$, $b=1$, $r=0$

So what is integrability? When does it arise? When is a given system
nonintegrable? What distinguishes nearintegrable ones and chaotic
systems and so on? These are some of the paramount questions which arises
as far as nonlinear systems are concerned. Systematic answers will pave the
way for a meaningful understanding of nonlinear systems in general and the
role of nonlinearity in particular.

As far as integrable systems are concerned the earlier discussions seem to
point out at least the following broad definitions:

1) \underline {Integrablity in the complex plane}:
Integrable - integrated with sufficient number of arbitrary constants or
functions; nonintegrable - proven not to be integrable. This loose definition
can be related to the existence of single-valued, analytic solutions a concept
originally advocated by Fuchs, Kovalevskaya, Painlev\'{e} and others, thereby
leading to the notion of ``integrability in the complex plane" and to the
Painlev\'{e} property mentioned in the previous section. For real valued
coordinates, this can lead to integration methods such as B\"{a}cklund
transformations, Hirota's bilinearization and ultimately Lax pair and inverse
scattering analysis to completely solve the Cauchy initial value problem.

2) \underline {Integrability - Existence of integrals of motion}:
One looks for sufficient number of single - valued, analytic integrals of
motion. For example, $N$ integrals for Hamiltonian systems with $N$ degrees
of freedom, involutive and functionally independent. Then the equation of
motion can be integrated by quadratures, leading to Liouville integrability.
Such a possibility leads to strong association with symmetries, generalized
Lie symmetries and Lie-B\"{a}cklund symmetries.

3) \underline {Integrability: Existence and uniquness of solutions}:
Mathematicians often call complete integrability as related to the existence
and uniqueness of solutions.

How are all these and other possible such concepts interrelated? Which is the
ultimate definition of integrability? In each of these definitions there are
pitfalls and one might construct some counterexamples, even if they are
pathological in nature. Then can one construct algorithmic ways of isolating
integrable nonlinear systems (whatever it ultimately means)
 and then analyse their dynamics systematically?

At least the definition of integrability in the complex plane seems to offer
such a possibility, however unsatisfactory the present status of Painlev\'{e}
analysis method is. Its applicability seems to be wide: it successfully
isolates integrable cases in nonlinear difference equations, nonlinear odes
(both of dissipative and Hamiltonian type), nonlinear pdes in (1+1), (2+1),
nonlocal, integro-differential equations and so on. It also captures other
integrability properties such as linearization, bilinearization, B\"{a}cklund
transformations and so on. However there are many unanswered or partially
answered questions, whose understanding can dramatically alter our
understanding of nonlinear systems.

1. Why does the method is successful in isolating integrable cases?
Integrability is something which we relate to real (space-) time dynamics.
Why should the properties in the complex plane/manifold determine the real
time behaviour?

2. Why is that certain type of singularities are bad while others are
admissible? Movable pdes of finite order are admissible but movable branch
points and essential singularities are associated with nonintegrability
and chaos? Then the P-property is defined to within a transformation. Which ones
are allowed?

3. Why are fixed singularities of all type including essential singularities
are allowed while movable critical singularities are not allowed? If
denseness of branching is to be a criterion why branching around fixed
singular points are allowed?

4. In nonintegrable cases, one observes that it is essential in typical
cases to develop the solutions as double infinite series around a movable
singular point or manifold. Can one extract real time behaviour from these
asymptotic forms? Do the complex patterns of singularities and apparent
fractal structure have any connetion with real time behaviour? [Bountis, 
1992]. Typical examples of Duffing oscillator and Duffing-van der Pol 
oscillator are given in Figs.13-16(see also Lakshmanan \& Murali[1996]).

5. Can any criterion for the denseness of branching be given? It may be
possible to develop the `PolyPainlev\'{e} test' proposed by Martin Kruskal
[Kruskal \& Clarkson, 1992] so as to understand the effect of denseness of
branching.

6. What is the connection between the nature of solutions around the
singular points/ manifolds with the integrals of motion and existence of
solutions?

It appears that in the next few decades determined efforts to understand
integrability and nonintegrability aspects along these lines can throw much
light on the nature of nonlinear systems, which then will lead to
algorithmic handles to deal with such systems in a general sense. This
will open up many paths to analyse nonlinear physical systems under very
many new circumstances.  Thus singularity structure may turn out to be the key
to unlock nonlinear dynamical systems in the next few decades.  

\subsection{Nonlinear excitations in higher spatial dimensions}
In the earlier section 4.8 we noted that richer physical structures can arise in
(2+1) dimensions. While in (1+1) dimension we have seen the
possibility that both localized coherent structures as well as chaotic
structures exist, one would like to know whether these excitations survive
in higher dimensions too and whether there are new elementary excitations and
new phenomena lurking in higher spatial dimensions. Unfortunately the natural
physical extensions of (1+1) dimensional soliton equations such as
sine-Gordon, nonlinear Schr\"{o}dinger or Heisenberg ferromagnetic spin
equations of the form
$$ \frac {\partial^2 \phi}{\partial t^2} -\nabla^2 \phi+m^2 \sin \phi =0,
\eqno (121) $$
$$ iq_t+\nabla^2q+|q|^2q=0, \eqno (122) $$
$$ \overrightarrow {S_t}=\vec S \times \nabla^2 \vec S, \; \vec {S}^2=1,
\eqno (123) $$
etc., where $\overrightarrow{\nabla}^2$ is the 2-dimensional or 3-dimensional
Laplacian, do not
seem to possess straightforward extensions of solitonlike exponential
localized structures. In fact, it is expected that the solutions of these
equations even might develop singularities, sometimes called collapse. These
equations sometimes may possess under special geometries interesting classes
of particular solutions, like time-independent spherically symmetric, axially
symmetric, instanton, vortex, monopole, hedgehog and so on solutions
(Rajaraman [1980]; Makhankov etal [1993]). But the nature of their general
solutions is simply not known.

On the other hand, at least in (2+1) dimensions, introduction of additional
nonlocal terms or effectively additional scalar fields giving rise to
boundary effects can offset the tendency for leakage of energy in the second
spatial direction so as to make the system admit exponentially localized
solutions, namely dromions, in addition to algebraically decaying lumps
(mentioned in Sec.4). The K-P, D-S and Ishimori equations and their typical
solutions are given in Sec.4, Table II. In additon, we have some of the other
interesting (2+1) dimensional equations, each one with nonlocal terms,
admitting exponentially localized solutions (see Radha \& Lakshmanan [1995]).

\noindent 1) \underline {(2+1) dimensional KdV}
$$ u_t+u_{\xi\xi\xi}=3(u \partial_{\eta}^{-1}u_{\xi})_{\xi} . \eqno (124) $$

\noindent 2) \underline {Niznik-Novikov-Velesov equation}
$$u_t+u_{\xi\xi\xi}+u_{\eta\eta\eta}+au_{\xi}+bu_{\eta}=
3(u \partial_{\eta}^{-1}u_{\xi})_{\xi}+
3(u \partial_{\xi}^{-1}u_{\eta})_{\eta} . \eqno (125)$$

\noindent 3) \underline {Generalized NLS}
$$ iq_t=q_{xy}+Vq, \eqno (126a) $$
$$ V_x=2\partial_y|q|^2. \eqno (126b) $$

\noindent 4) \underline {(2+1) dimensional sG}
$$ \theta_{\xi\eta t}+\frac{1}{2}\theta_{\eta}\rho_{\xi}
+\frac{1}{2}\theta_{\xi}\rho_{\eta}=0, \eqno (127a) $$
$$ \rho_{\xi\eta}=\frac{1}{2}(\theta_{\xi}\rho_{\eta})_t. \eqno (127b) $$

\noindent 5) \underline {(2+1) dimensional simplest scalar equation}
$$iq_t+q_{xx}-2\lambda q \int_{-\infty}^y|q|_x^2 dy'=0, \;
\lambda=\pm1. \eqno (128) $$

\noindent 6) \underline {2D long dispersive wave equation}
$$ \lambda q_t+q_{xx}-2qv=0, $$
$$ \lambda r_t+r_{xx}+2rv=0, $$
$$ (qr)_x=v_{\eta}, \; \partial_{\eta}=\partial_x-\partial_y. \eqno (129) $$

So if boundary contributions are important for localized solutions to exist,
then what about the nature of the excitations in the physically important
systems such as (121-123) given above. Numerical investigations are time
consuming and require considerable effort. For example, for nonlinear
$\sigma$-model Zakrzewski and coworkers [1995] have obtained interesting
numerical results, including scattering, elastic and inelastic collisions.
Typical results are given in Fig.(16).
The following problems need urgent attention,
which might point towards new vistas in nonlinear physics.

1. What are all the possible stable structures (special solutions) in
(2+1) and (3+1) dimensions? How stable are they? What are their
collision properties? If unstable, are they metastable? If not do they
give rise to new spatio-temporal pattern? What is the effect of external
forces including damping?

2. Can one develop techniques to solve the Cauchy initial value problem
of physically important (2+1) dimensional extensions of (1+1)
dimensional soliton equations? Or can one obtain enough informations
about the nature of general excitations through numerical analysis? How
can the numerics be simplified to tackle such problems?

3. Is it possible to perceive something similar to FPU experiments in
(2+1) and (3+1) dimensions? What new phenomena are in store here? Can
actual analog simulations be made with suitable miniaturization of
electronic circuits so that these (2+1) and (3+1) dimensional systems
can be analysed systematically?

4. Is it possible to extend the inverse scattering formulation to (3+1)
dimensional systems also as in the case of (2+1) dimensional evolution
equations? The main difficulty seems to arise in the inverse analysis
due to certain nonuniqueness arising from constraints on the scattering
data (see for example, Ablowitz \& Clarkson [1991]).

It is very certain that the future of nonlinear physics will be much
concentrated around such higher dimensional nonlinear systems, where new
understandings and applications will arise in large numbers.
A long term sustained numerical and theoretical analysis of (2+1) and (3+1)
dimensional nonlinear evolution equations both for finite and continuous
degrees of 
freedom will be one of the major tasks for several decades 
to come which can throw
open many new nonlinear phenomena. Also one might consider discretization and
analog simulation of these systems, to which nonlinear electronics
community can contribute much.

\subsection{Nonintegrable systems, spatio-temporal patterns and chaos}
In the earlier sections we considered integrable nonlinear systems. However
these are far fewer in number. Most natural systems are nonintegrable:
however many of them may be considered as \underline {perturbations of
integrable nonlinear} systems. Examples: condensed matter systems including
magnetic, electronic and lattice systems, optoelectronic systems,
hydrodynamical sytems and so on. The perturbing forces could be space-time
inhomogeneities and modulations, external forces of different origins, damping
and dissipative as well as diffusive forces and so on. Thus
it is imperative to study the effect of these various additional forces with
reference to the basic nonlinear excitations of integrable systems.

Such an analysis needs to consider the different length scales of the
perturbation (both space and time) with respect to the nonlinear excitations
of the unperturbed case [Scharf, 1995; Kivshar \& Spatcheck, 1995]. Depending
on such scales, the original entities might survive albeit necessary
deformations or may undergo chaotic or complex motions or deformations may
give rise to interesting spatio-temporal patterns. Some preliminary studies
on such soliton perturbations are available in the literature (see for
example, Scharf[1995], Kivshar \& Spatcheck [1995]). In fig.17, a typical
soliton perturbation in the case of the nonlinear Schr\"{o}dinger equation
$$iu_t+u_{xx}+2u|u|^2=\epsilon u\cos (kx) \eqno (130) $$
is seen to give rise to a spatio-temporal pattern.

Detailed classification of the types of perturbations and the resulting
coherent and chaotic structures and spatio-temporal patterns can be used
as dictionaries to explain different physical situations in condensed
matter, fluid dynamics, plasma physics, magnetism, atomospheric physics
and so on. Further such studies in (2+1) dimensional systems, wherein
any stable entity when perturbed by additional weak forces can lead to
exciting new structures corresponding to realistic world description. A
concerted effort through analytical and numerical investigations to
tackle the nonintegrable systems using integrable structures portends to
provide rich dividends in the next century.

\subsection{Micromagnetics and magnetoelectronics}
Micromagnetics is the subject which is concerned with the study of
detailed magnetization configurations and the magnetization reversal
process in magenetic materials (Brown Jr. [1963]). Particularly it
encompasses the study of ferromagnets and ferromagnetic thin films
(used in magnetic thin film sensors and devices). The theory considers
the ferromagnetic free energy in the ferromagnectic material to consist of

i) the ferromagnetic exchange energy,

ii) the magnetic anisotropy energy,

iii) the magnetoelastic energy,

iv) the magnetostatic energy and

v) the magnetic potential energy due to external magnetic fields.

The magnetization orientation $\overrightarrow M(\vec r,t)$
follows the Landau-Lifshitz-Gilbert equation (Landau \& Lifshitz [1935];
Lakshmanan \& Nakamura [1985])
$$ \frac{\partial {\overrightarrow M}}{\partial t}=-\gamma \overrightarrow
M \times \overrightarrow F_{eff}-\frac{\lambda}{M}\overrightarrow M \times
(\overrightarrow M \times \overrightarrow F_{eff} ), \eqno (131) $$
where $\gamma$ is the gyromagnetic ratio and $\lambda$ is the damping
constant and $\overrightarrow F_{eff}$
is the effective magnetic field. The first term
in the equation describes the the gyromagnetic motion (precession of
$\overrightarrow M$ about $\overrightarrow F_{eff}$), and the second
describes the rotation of the effective field. Note that $|\overrightarrow
M|=$constant. The Heisenberg ferromagnet equations discussed earlier in
Sec.4.7.1 are then essentially special cases of the above Landau-Lifshitz
equation when the damping vanishes and $\overrightarrow F_{eff}$ takes
special forms. The complex magnetization patterns and the detailed spin
structures within the domain boundaries are obtained by solving eq.(131). The
structures so obtianed  can then be used for different applications (See
also the special issue on ``Magnetoelectronics"-Physics Today, April 1995).

\noindent 1. \underline {Magnetoresistive recording}

Over a century now, for magnetic recording most systems have used an
inductive head for writing and reading which employ coils to both induce
a magnetic field (write mode) and sense a recorded area (read mode). Recently,
a more powerful reading head called the \underline{magnetoresistance (MR)
head}, has been introduced into disk products which employs a sensor whose
resistance changes in the presence of a magnetic field. Its performance gain
has enhanced the density of storage by upto 50\% commercial conversion to
MR heads is only just starting.

\noindent 2. \underline{Magneto-optical recording}

Again instead of using the conventional inductive head for recording,
optical pulses or lasers have become usage. Optical recording is expected
to increase in capacity and transfer rate by a factor about 20 over the
next decade. Rewritable systems will be based largely on magneto optical 
technologies that exploit the smaller mark sizes made possible by new
short wavelength lasers.

In the above two models, it is mainly the interaction between the
magnetization of the medium and the electrical field of the head in the case
of magnetoresistive head and the electromagnetic (optic or laser) field in
the case of magneto optical head that play important roles. For efficient
storage, excitation of the magnetization of the medium (which may be due to
thermal or due to other external disturbances) if any should be localized.
Theoretically speaking, these different magnetic interactions can be
accomodated in appropreate spin Hamiltonian models.

\noindent 3. \underline{Magnetic films for better recording}

The fundamental magnetization process in thin films can be characterized
by the formation, motion and annihilation of magnetization vortices. When a
sufficiently strong external magnetic field is applied, magnetization reversal
takes place and these vortices move across the film. If the intergranular
exchange coupling in magnetic films is large the size of the vortices will
be larger and travel more freely over larger distances. Thus the intergranular
exchange coupling has a significant impact on the properties of recorded bits
in thin films because in the case of large vortices the recording noise is
large and in the case of low vortices the noise is low.

Vortices form an interesting class of solutions to multidimensional nonlinear
evolution equations. Hence here also solving higher dimensional Landau-
Lifshitz equations with intergranular exchange coupling and interaction with
large external fields for vortex like solutions is an important task for
future.

\noindent 4. \underline{Study of single domain magnets}

The behaviour of indivdual magnetic domains has become important for technology.
The problem of media noise which is one of the fundamental present day
problems of magnetic storage can be avoided if we use individual magnetic
domains to store each bit. The study of single domain magnets (mesoscopic
magnets) has accelerated the development of new theoretical approachs to
magnetic dynamics. Thus for the next few years attention has to be paid
how quantum mechanical effects influence the properties of all these small
systems.

\subsection{Optical soliton based communication: perspectives and
potentialities}
Soliton based optical-fiber communication is imminent, as we have noted
in Sec.4.7.2. The experimental works of Mollenauer and co-workers [1990]
has clearly demonstrated the successful soliton transmission over more than
10,000kms in a dispersion-shifted fiber. With such an exciting possibility,
it is important to study further technical effects related to soliton 
propagation in optical fibers. We mention here a few of them.

1. The assumption of instantaneous nonlinear response amounts to neglecting
the contriibution of molecular vibrations in the higher-order nonlinear
effect. In general, both electrons and nuclei respond to the optical field in
a nonlinear manner. For silica fiber, in the femto-second region, the
higher-order nonlinear effects (higher-order nonlinear dispersion effect,
self-induced Raman scattering effect)  become important.
Further in the near zero group velocity dispersion region, higher-order 
dispersion term becomes essential. Typically the NLS equation (83) gets
 modified to the form
$$ iE_z-\frac{\kappa''}{2}E_{\tau\tau}+\frac{n_2\omega}{c}|E|^2E -
\frac{i\kappa'''}{6}E_{\tau\tau\tau}+i\gamma(|E|^2E)_{\tau}+
i\gamma_s(|E|^2)_{\tau}E=0, \eqno (132) $$
where  $\kappa'''=\frac{\displaystyle \partial^3 \displaystyle \kappa}
{\displaystyle \partial \displaystyle \omega^3}$ describes
third order dispersion and $\gamma$ describes nonlinear dispersion and
$\gamma_s$ stands for self-induced Raman scattering effect. Much works need
to be done in the analysis of the above type of equations. Similarly for
erbium doped fibers, which are quite useful from the point of view of self-amplification,
one has to analyse the full soliton dynamics of Maxwell-Bloch
equations.

2. In most experimental situations of propagation of light pulses in nonlinear
medium, pulsed laser is used as a source for excitation, especially when
high-power operation is involved. In these circumstances, a beam of
light will experience both diffraction and dispersion, in additon to the
self-focussing (defocussing) and self-phase modulation that results from
the nonlinearity. The corresponding evolution equation is the higher
dimensional NLS equation of the form given in eq.(132). What kind of pulses
do such equations admit which can be considered as 
non-diffractive and non-dispersive pulses of experimental relevence? Can one
have a stable light bullet (soliton) which can be used experimentally?

3. Since SiO$_2$ is a symmetric molecule, second order nonlinear effect
vanishes. Neverthless the electric-quadrupole and magnetic-dipole
moments can generate weak second-order nonlinear effects.  Defect or
color centers inside the fiber core can also contribute to
second-harmonic generations under certain conditions. What is the role of
such second-order nonlinear effect on the optical soliton prpagation?

4.  In many circumstances, a more complete description of the
propagation would rather involve an interaction between two (or more)
coupled modes.  For example, birefringence will give rise to two
nondegenerate polarization modes.  The coupling could also be between the
modes of two optical guides as in dual-core fiber.  Coupled nonlinear
Schr\"{o}dinger family of equations are used to study these phenomena. 
Investigation of these types of models is a very important current and
future direction in this field as pulse transmission devices that rely
on coupling between fibers(e.g, switches, directional couplers) are
often components of optical fiber communication systems.  In addtion,
investigation of nonlinear processes in coupled optical waveguides will
aid in the design of various optical computers and sensor elements.

In short, there are very many pressing problems to be investigated on the
nonlinear Schr\"{o}dinger family of equations, depending upon the physical
situation in which the light propagation in nonlinear fibers takes
place. Sustained investigations can throw much insight into the basic 
phenomenon, besides helping the actual soliton based information
technology to develop into a favoured method of information transmission
in the coming decades.

\section { Conclusions and Outlook}
Nonlinear physics ( and nonlinear science) has come a long way from a
position of insignificance to a central stage in physics and even in science
as a whole.  While the pace of such a development was rather slow,
an eventful golden era which ensued during the period 1950-70 saw the
stream-rolling of the field into an interdisciplinary topic of great
relevance of scientific endeavour.  We have tried to present here a rather
personal perspective of some of these developments and the outstanding tasks
urgently need to be carried out in the forthcoming decades and the possible
dividends they can bring in.  Of course these forecasts are based on the
speaker's own understanding and knowledge of the specific areas he is familiar
with.  Though the motivation has been to cover both integrable and chaotic
nonlinear systems, specific emphasis was given to integrable systems and
associated coherent structures, since the topic of chaos is covered in many
of the other lectures of the meeting.

Of course it is clear that the future developments and directions we indicate
here depend mostly on the present status of the concerned topics.  But one
is also strongly aware of the fact that path breaking new ideas and directions
can arise from nowhere suddenly and dramatically and without forewarnings
at anytime in the future.  The past and immediate present developments
described in this article definitely substantiate such unforeseen possibilities.
As stressed earlier there was no field called nonlinear science or nonlinear
physics fifty years ago and we cannot naturally foresee exactly how the
field would have transformed in the next fifty years.

However, we have tried to stress that much new physics can come out by 1)
clearly understanding the concepts of integrability and nonintegrability from
a unified point of view, 2) by analysing nonlinear structures in (2+1) and
(3+1) dimensional spaces, which are more realistic, and 3) through indepth
analysis of the effect of perturbation on integrable nonlinear systems and
analysis of other nonintegrable systems and classifying the types of novel
spatio-temporal structures which might arise.  We have also tried to point
out some of the tasks and potentialities in certain emerging technology
oriented topics such as magneto-electronics, optical soliton based
communications and so on, which are the off-shoots of progress at the
fundamental level.

There are numerous important topics which we have not touched upon or
discussed their future developments here, including such topics as
nonlinearities in plasma physics, acoustics, biological physics, 
many areas of condensed matter physics,
astrophysics, gravitational theory, detailed quantum aspects and so on.
Probably experts in these topics will cover such areas in future.  Similarly
the quest towards the ultimate theory of matter in particle physics,
whichever form it may take, will ultimately be a nonlinear one.  There is no
doubt that nonlinearity will rule the world for many more years to come
and there is scope for everybody to try his hand in the field for a better
understanding of Nature.

\vskip 7pt
\noindent {\bf {Acknowledgements:}}
\par
This work forms part of a Department of Science and Technology, Government 
of India research project.

\newpage
\parindent 0cm
\subsection*{Figure Captions}

{Fig.1} {The circuit diagram of the van del Pol oscillator}

{Fig.2} Subharmonics of the van der Pol oscillaor

{Fig.3} Energy sharing in the FPU anharmonic lattice between the various
modes

{Fig.4} Zabusky-Kruskal numerical analysis of the KdV equation: birth of
solitons

{Fig.5} Two-soliton scattering in the KdV equation

{Fig.6} Schematic digram of the IST method

{Fig 7} Hamiltonian chaos in Henon-Heils system: Poincar\'{e} surface
of section for different energies

{Fig.8} A chaotic attractor of the Lorenz system

{Fig.9} Methods for solutions of soliton equations

{Fig.10} A line soliton solution for the KP-I equation

{Fig.11} Lump-soliton solution of KP-I equation

{Fig.12} Dromion solution of the DS-I equation

{Fig.13} Square lattice of pole singularities in the complex t-plane of
a free undamped  Duffing oscillator ($\ddot x +
\omega_0^2 x +\beta x^3 = 0, \; \omega_0^2 = 1,\; \beta = 5$)

{Fig.14} Singularity distribution in the complex t-domain of the damped
Duffing oscillator equation ($\ddot x
+0.1 \dot x + \omega_0^2 x + \beta x ^3= 0, \; \omega_0^2 = 1,\; \beta = 5$)

{Fig.15} Singularity clustering for the driven (chaotic) Duffing oscillator
($\ddot x+ x + 5x^3= f\cos\omega t$)

{Fig.16} Scattering of two skyrmions [Piete \& Zakrzewski, 1995]

{Fig.17} Complicated spatio-temporal behaviour of the soliton in the
perturbed NLS equation [Scharf, 1995]

\end{document}